%% file: main.tex
\documentclass[journal=jpclcd,manuscript=letter,layout=twocolumn]{achemso}

\usepackage[version=3]{mhchem}
\usepackage{xcolor}
\usepackage{tikz}
\usetikzlibrary{decorations.pathreplacing}
\usetikzlibrary{arrows,positioning} 
\usetikzlibrary {arrows.meta} 
\usetikzlibrary{pgfplots.groupplots}
\usepackage{pgfplots}
\pgfplotsset{compat=1.18}
\usepackage{subcaption}
\usepackage[caption=false]{subfig}
\usepackage{physics}
\usepackage{lipsum}

\makeatletter
\newcommand{\citenumber}[1]{%
    \begingroup
    \let\@cite\NAT@citenum
    \>\cite{#1}%
    \endgroup
}
\makeatother

\author{Lizeth Franco}
\affiliation{Departamento de F\'isica y Qu\'imica Te\'orica, Facultad de Qu\'imica, Universidad Nacional Aut\'onoma de M\'exico, M\'exico City, C.P. 04510, M\'exico}
\author{Iván A. Bonfil-Rivera}
\affiliation{Departamento de F\'isica y Qu\'imica Te\'orica, Facultad de Qu\'imica, Universidad Nacional Aut\'onoma de M\'exico, M\'exico City, C.P. 04510, M\'exico}
\author{Juan Felipe Huan Lew-Yee}
\email{felipe.lew.yee@dipc.org}
\affiliation{Donostia International Physics Center (DIPC), 20018 Donostia, Spain.}
\author{Mario Piris}
\email{mario.piris@ehu.eus}
\affiliation{Donostia International Physics Center (DIPC), 20018 Donostia, Spain.}
\alsoaffiliation{Kimika Fakultatea, Euskal Herriko Unibertsitatea (UPV/EHU), 20018 Donostia, Spain.}
\alsoaffiliation{IKERBASQUE, Basque Foundation for Science, 48013 Bilbao, Spain.}
\author{Jorge M. del Campo}
\email{jmdelc@unam.mx}
\affiliation{Departamento de F\'isica y Qu\'imica Te\'orica, Facultad de Qu\'imica, Universidad Nacional Aut\'onoma de M\'exico, M\'exico City, C.P. 04510, M\'exico}
\author{Rodrigo A. Vargas-Hernández}
\affiliation[Mc]{Department of Chemistry and Chemical Biology, McMaster University, Hamilton, ON, Canada}
\alsoaffiliation[BIMR]{Brockhouse Institute for Materials Research, McMaster University, Hamilton, ON, Canada}
\email{vargashr@mcmaster.ca}
\title[An \textsf{achemso} demo]
  {Softmax parameterization of the occupation numbers for natural orbital functionals based on electron pairing approaches}
\begin{document}
\makeatletter
\setlength\acs@tocentry@height{2in}
\setlength\acs@tocentry@width{2in}
\makeatother
\begin{tocentry}
\includegraphics[scale=0.12]{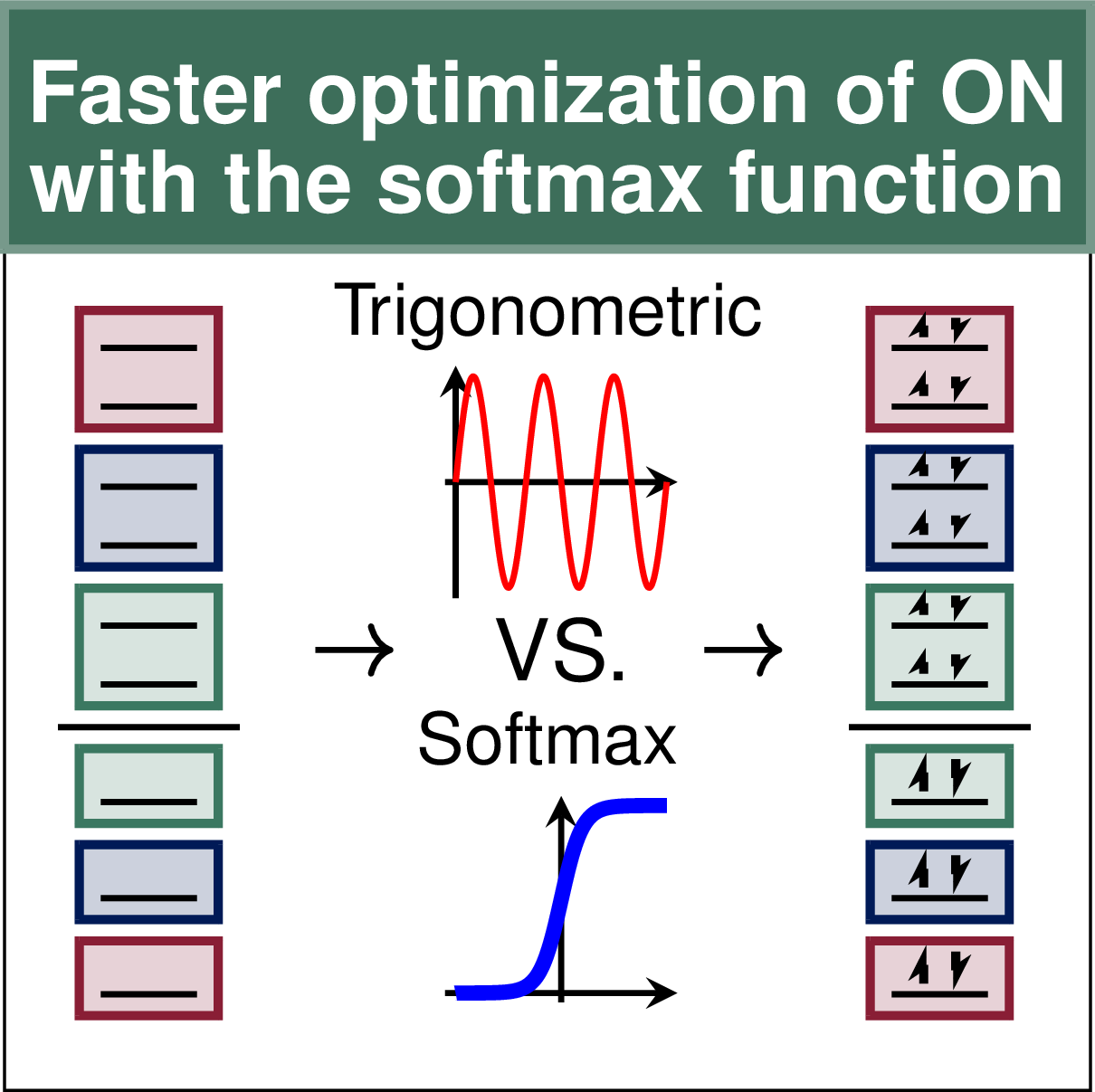}
\end{tocentry}

\begin{abstract}
  Within the framework of natural orbital functional theory, having a convenient representation of the occupation numbers and orbitals becomes critical for the computational performance of the calculations. Recognizing this, we propose an innovative parametrization of the occupation numbers that takes advantage of the electron-pairing approach used in Piris natural orbital functionals through the adoption of the softmax function, a pivotal component in modern deep-learning models. Our approach not only ensures adherence to the N-representability of the first-order reduced density matrix (1RDM) but also significantly enhances the computational efficiency of 1RDM functional theory calculations. The effectiveness of this alternative parameterization approach was assessed using the W4-17-MR molecular set, which demonstrated faster and more robust convergence compared to previous implementations.
\end{abstract}

\section{Introduction}
Reduced-density-matrix-based methods promise higher accuracy compared to density functional approximations while being computationally less demanding than wavefunction methods.\cite{Mazziotti2006-mk} In particular, one-electron reduced density matrix functional theory (1RDMFT) is rooted in Gilbert's extension~\cite{Gilbert1975} of the Hohenberg-Kohn theorem,\cite{Hohenberg1964-lx} further developed through the efforts of Levy~\cite{Levy1979-dw} and Valone,\cite{Valone1980-yh} who demonstrated the existence of an exact, yet undiscovered, functional of the first-order reduced density matrix (1RDM). On the other hand, Coleman's seminal work~\cite{Coleman1963-ur} underscored the importance of constraining the 1RDM to be N-representable, which can be achieved by restricting its eigenvalues within the zero to one interval, and ensuring that its sum equals the total number of electrons~\cite{Lowdin1955-qk}. In the natural orbital (NO) representation, the 1RDM becomes diagonal, with the occupation numbers (ONs) being its eigenvalues. The conditions for the ONs can be cast as follows:
\begin{equation}
    0 \leq  n_i \leq 1, \text{  and    } \sum_i n_i = \mathrm{N},\label{eqn:lowdin_const}    
\end{equation}
where $n_i$ is the ON of the $i^{th}$-NO, $\mathrm{N}$ is the total number of electrons, and the sum goes over all spin natural orbitals. This led to the development of functionals based on NOs and their ONs, transforming the 1RDMFT into a natural orbital functional theory (NOFT).\cite{Muller1984-ji,Goedecker1998-pj,Gritsenko2005-rl,Piris2006-kn,Lathiotakis2008-ph} At this point, it is worth acknowledging that the investigation of the 1RDM functional remains an active research field with promising applications by itself,\cite{S18, SS19, SP21, LCLS22, CSS21, CLLPPS23, LCS23} while its advances can directly benefit the development of NOFT. On the other hand, we also advise the reader to consult a recent review article to obtain a comprehensive historical account of the formulation and development of NOFT.\cite{Piris2024}

Routine NOF calculations frequently involve simultaneous or alternate optimization of NOs and their ONs. For this reason, the development of effective optimization schemes plays a critical role in facilitating the widespread adoption of NOFT in quantum chemistry simulations. Orbital optimization can be performed using iterative diagonalization,\cite{Piris2009-jo} and orbital rotations,~\cite{Herbert2003,Elayan2022-nb} while several approaches have been proposed to perform ON optimization in an unconstrained manner while preserving Eq.~(\ref{eqn:lowdin_const}) conditions.\cite{Baldsiefen2013-ob,Piris1999-mn,Piris2021-xo} 
For example, Yao \emph{et al.} \cite{Yao2021-gm,Yao2022-sa} recently proposed an explicit by implicit (EBI)~\cite{Yao2021-gm,Yao2022-sa} representation of ONs, stimulating various creative approaches to deal with convergence.\cite{cartier2024exploiting,yao2024enhancing} The EBI method is notable for its provision of a conducive surface for auxiliary optimization parameters, representing a significant advancement in ON optimization.

In line with various electronic structure methods, RDMs have been integrated into machine learning frameworks, for example, to evaluate the energies and atomic forces of small molecules.\cite{Shao2023} In addition, convolutional neural networks have been used in the prediction of 2RDM's eigenvalues.\cite{Sager-Smith2022,Grier2023}. The present work is directly inspired by these developments. However, we took a different approach, aiming to harness the softmax activation function utilized in machine learning models~\cite{mlphysbook:2023,kunc2024}, to parameterize ONs in a manner that satisfies Eq.~(\ref{eqn:lowdin_const}). Our approach is rooted in the belief that the integration of machine learning techniques with NOFT can lead to significant improvements in solving convergence issues, thereby enhancing its applicability and effectiveness in quantum chemistry RDM-based simulations.\vspace{-0.25cm}

\section{Methods}
Recent advances in the field of NOFT have been marked by the development of electron-pairing-based Piris NOFs (PNOFs)~\cite{Piris2011-qn,Piris2013-qj,Piris2014-bj,Piris2017-go}. PNOFs have shown remarkable performance in addressing the charge delocalization error,\cite{Lew-Yee2022-te,Huan-Lew-Yee2023-ha} and handling challenging multireference problems.\cite{Mitxelena2022,MERCERO2023229,Lew-Yee2022-iron,Franco2023} The latest addition to this family is the Global NOF~\cite{Piris2021-sv}, which aims to strike a balance between static and dynamic electron correlation. In particular, it represents a promising approach with manageable computational demands~\cite{Lew-Yee2021-bb}, highlighting the importance of ongoing development and motivating the exploration of new convergence schemes that can benefit these methodologies.

In the following, we consider $\mathrm{N_{I}}$ unpaired electrons, which determine the total spin of the system, $S$. The remaining electrons, $\mathrm{N_{II}} = \mathrm{N-N_{I}}$, form pairs with opposite spins, resulting in a net spin of zero for $\mathrm{N_{II}}$ electrons combined. For the equally mixed state of highest multiplicity, defined by $2S+1=\mathrm{N_{I}}+1$,\cite{Piris2019} the expected value of $\hat{S}_{z}$ is zero. Consequently, restricted spin theory is applied: $\varphi_p^{\alpha} \left(\mathbf{r}\right ) = \varphi_p^{\beta} \left(\mathbf{r}\right) = \varphi_p \left(\mathbf{r}\right), n_p^ {\alpha} =n_p^{\beta}=n_p$. 

\begin{figure}[htb]
    \centering
    \input{Figures/pairing-scheme-F}
    \caption{Splitting of the orbital space $\Omega$ into subspaces for a system with nine electrons in a doublet spin state ($\mathrm{N_{I}}$=1, 2$S$+1=2). In this example, one orbital make up the subspace $\Omega_{\mathrm{I}}$=$\Omega_{5}$, whereas eight electrons ($\mathrm{N_{II}}=8$) distributed in four subspaces $\left\{ \Omega_{1},\Omega_{2},\Omega_{3},\Omega_{4}\right\} $ make up the subspace $\Omega_{\mathrm{II}}$. Note that $\mathrm{N}_{g}$=2 for all subspaces $\Omega{}_{g}\in\Omega_{\mathrm{II}}$. The arrows depict the values of the ensemble occupancies, alpha ($\downarrow$) or beta ($\uparrow$), in each orbital $\left|p\right\rangle$.}  
    \label{fig:pnof-orbital-pairing-F}
\end{figure}
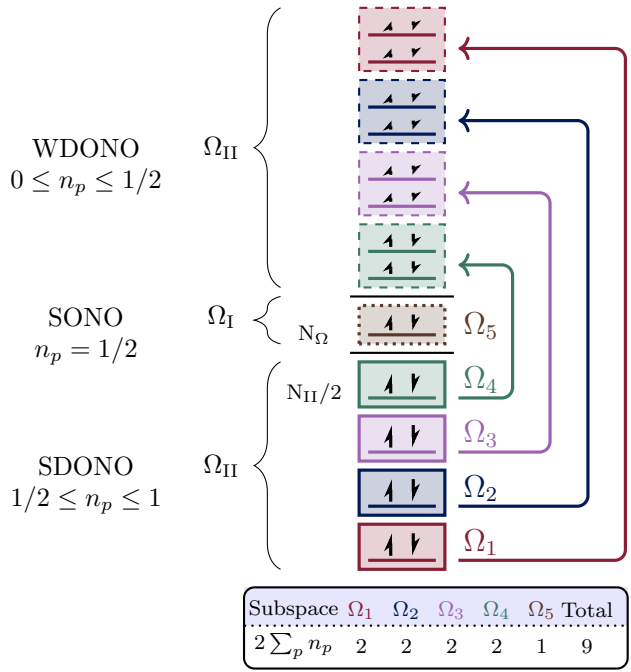

The orbital space $\Omega$ is divided into disjoint subspaces $\Omega_g$=$\left\{ \varphi_g, \varphi_1, \varphi_2,...,\varphi_{\mathrm{N}_{g}} \right\}$, which leads to explicit terms for intra and inter-pair electron correlation~\cite{Piris2018-rr}. The value of $\mathrm{N}_{g}$ may vary for each subspace, with its maximum determined by the basis set. An illustration of $\Omega$ can be seen in Fig. \ref{fig:pnof-orbital-pairing-F}, depicting the case of a doublet system with nine electrons and $\mathrm{N}_{g}=2$. There are four strongly double-occupied NOs (SDONO), each coupled with two weakly double-occupied NOs (WDONO), comprising the subspace $\Omega_{\mathrm{II}}$ distributed in four subspaces ${\left\{\Omega_1,\Omega_2,\Omega_3,\Omega_4\right\}}$. Taking into account the spin, the total occupancy for a given subspace in $\Omega_{\mathrm{II}}$ is 2, which is reflected in the following pairing condition sum rule for the orbitals of a given spin,
\begin{equation}
\sum_{p\in\Omega_{g}}n_{p}=n_{g}+\sum_{i=1}^{\mathrm{N}_{g}}n_{p_{i}}=1; \>\> g=1,\cdots,\frac{\mathrm{N_{II}}}{2} \label{eq:sum1}.
\end{equation}
In addition, the single occupied NO (SONO) in $\Omega_5$ belongs to $\Omega_\mathrm{I}$ and is responsible for the doublet state. It is important to emphasize that each orbital in $\Omega_{\mathrm{I}}$ accommodates a single electron ($n_p=\tfrac{1}{2}$), but the specific spin state, whether $\alpha$ or $\beta$, is unknown. 

The electron-pairing strategy effectively converts the normalization condition into multiple sum rules designed for electron pairs, as shown in Eq.~(\ref{eq:sum1}), ensuring the inherent fulfillment of Eq.~(\ref{eqn:lowdin_const}),
\begin{equation}
2\sum_{p\in\Omega}n_{p}=2\left (\sum_{p\in\Omega_{\mathrm{I}}}+\sum_{p\in\Omega_{\mathrm{II}}}\right)n_{p}=\mathrm{N_{I}}+\mathrm{N_{II}}=\mathrm{N}.\label{norm}
\end{equation}

In the following, we review the current state-of-the-art of occupation number parameterization by trigonometric functions in the context of electron-pairing NOFs, then, we present a parameterization based on the softmax function. For this work's context, the goal is to promote the use of the \texttt{softmax} function~\cite{softmax:1989} as an alternative parametrization of the ONs while leveraging the electron-pairing conditions [Eq.~(\ref{eq:sum1})] employed in PNOF.

\subsection{Trigonometric parameterization}
A notable aspect of the orbital pairing approach is that it directly enables unconstrained minimization of the ONs, which is particularly advantageous for optimization in NOFT. This has long been performed employing trigonometric functions.~\cite{Piris1999-mn} In the specific case at hand, the ONs can be represented by means of squared trigonometric functions. For a subspace $\Omega_g$, the occupation number $n_g$ of the SDONO is given by 
\begin{equation}
    n_g = \frac{1}{2} \left(1 + \cos^2 \gamma_g \right), \>\> g={1,\cdots,\frac{\mathrm{N_{II}}}{2}} ,
    \label{eq:n_sdono_gamma}
\end{equation}
while the occupation number of the WDONOs, $n_{p_i}$, are given by, 
{\scriptsize
\begin{eqnarray}
    \label{eq:n_wdono_gamma1}
    n_{p_1} \!\!\!\!&=&\!\!\! h_g \sin^2 \gamma_{p_1} ,\nonumber\\
    n_{p_2} \!\!\!\!&=&\!\!\! h_g \cos^2 \gamma_{p_1} \sin^2 \gamma_{p_2} ,\nonumber\\
    &\vdots& \nonumber\\
    n_{p_{i}} \!\!\!\!&=&\!\!\! h_g \cos^2 \gamma_{p_1} \cos^2 \gamma_{p_2} \cdots \cos^2 \gamma_{p_{i-1}} \sin^2 \gamma_{p_i},\nonumber\\
   &\vdots& \nonumber\\
   n_{p_\mathrm{N_{g}-1}} \!\!\!\!&=&\!\!\! h_g \cos^2 \gamma_{p_1} \cos^2 \gamma_{p_2} \cdots \cos^2 \gamma_{p_{N_{g-2}}} \sin^2 \gamma_{p_\mathrm{N_{g-1}}}, \nonumber\\    
   n_{p_\mathrm{N_{g}}} \!\!\!\!&=&\!\!\! h_g \cos^2 \gamma_{p_1} \cos^2 \gamma_{p_2} \cdots \cos^2 \gamma_{p_{N_{g-2}}} \cos^2 \gamma_{p_\mathrm{N_{g-1}}}, \nonumber\\
\end{eqnarray}}
where $h_g = 1 - n_g$. It can be easily shown that the pairing condition in Eq.~(\ref{eq:sum1}) is inherently guaranteed by the Pythagorean trigonometric identity~\cite{Piris2021-xo}.

Analytic gradients for the occupation numbers, with respect to the auxiliary variables, can be directly obtained as follows,

{\scriptsize
\begin{eqnarray}
    \frac{d n_{g}}{d \gamma_g} \!\!\!\!\!&=&\!\!\!\! - \frac{1}{2}\sin(2\gamma_g), \>\> g={1,\cdots,\frac{\mathrm{N_{II}}}{2}} ,\nonumber\\
    \frac{\partial n_{p_i}}{\partial \gamma_g} \!\!\!\!\!&=&\!\!\!\! - \frac{d n_g}{d \gamma_g} \cos^2 \gamma_{p_1} \cdots \cos^2 \gamma_{p_{i-1}} \sin^2 \gamma_{p_i}, \>\> p_i \in \Omega_g,\nonumber\\
    \frac{\partial n_{p_i}}{\partial \gamma_{p_j}} \!\!\!\!\!\!&=&\!\!\!\! - h_g \cos^2 \gamma_{p_1} \cdots \sin 2\gamma_{p_j} \cdots \cos^2 \gamma_{p_{i-1}} \sin^2 \gamma_{p_i}, \>\> j<i\neq \mathrm{N}_g ,\nonumber\\
    \frac{\partial n_{p_i}}{\partial \gamma_{p_i}} \!\!\!\!\!&=&\!\!\!\! h_g \cos^2 \gamma_{p_1} \cdots \cos^2 \gamma_{p_{i-1}} \sin 2\gamma_{p_i}, \>\> i \neq \mathrm{N}_g,\nonumber\\
    \frac{\partial n_{p_i}}{\partial \gamma_{p_j}} \!\!\!\!\!&=&\!\!\!\! - h_g \cos^2 \gamma_{p_1} \cdots \cos^2 \gamma_{p_{i-1}} \sin 2\gamma_{p_i}, \>\> j<i = \mathrm{N}_g.
\end{eqnarray}}

Overall, the parameterization effectively transforms the optimization of the ONs into a minimization of the unbounded auxiliary variables (${\gamma_p}$). However, since this parameterization involves the product of trigonometric functions in both the ONs and their gradients, minor deviations in $\gamma_p$ values can lead to inconvenient oscillatory propagation of the error, potentially causing numerical instabilities during the simulations. These deviations are amplified as the number of orbitals in a subspace increases because of the involvement of more products of trigonometric functions (Fig.~\ref{fig:trigonometric-softmax}). In practice, this parameterization performs well when there are few coupled NOs. It is worth acknowledging that this approach has greatly facilitated successful PNOF calculations across diverse complex systems, playing a significant role in its ongoing development.

\subsection{Softmax parameterization}
Higher accuracy is demanded in electronic structure calculations, so larger basis sets are required, resulting in more NOs for each subspace. Therefore, developing a more stable parameterization for the ONs becomes imperative. The softmax function can fulfill the pairing condition of Eq.~(\ref{eq:sum1}), providing a better way to take advantage of these properties with a function defined with simple elementary functions on all parts of the domain, making it easier to implement and evaluate. Moreover, this function is commonly used in deep learning models as it parameterizes a probability distribution over multiple classes~\cite{softmax:1989,mlphysbook:2023,kunc2024}, and was also used for the alchemy design of organic electronic materials~\cite{vargash:huxel:2023}.

Given a subspace $\Omega_g$ and using the softmax function, the ON can be parameterized as 
\begin{eqnarray}
    \label{eq:n_dono_softmax}
    n_{p} = \sigma(\gamma_p) = \frac{e^{\gamma_p}}{\sum_q e^{\gamma_{q}}}, 
\end{eqnarray}
where $p$ and $q$ are the indices of NOs, that is, they can be $\{g,\{p_i\}_1^{N_\text{g}}\}$, and $\gamma_p$ are auxiliary variables, also known as logits in the context of machine learning. The output of the softmax function is always normalized $\sum_p \sigma(\gamma_p) = 1$; therefore, it directly provides values for the ONs of a given subspace that fulfills Eq.~(\ref{eq:sum1}), as opposed to other machine learning activation functions~\cite{kunc2024}.
It is worth mentioning that the softmax function, also known as the softargmax or normalized exponential function \cite{softmax:1989}, is homologous to the Boltzmann distribution. Furthermore, the softmax function has closed-form gradients~\cite{softmax:1989,kunc2024}, as shown in Eq.~(\ref{eqn:softmax_grad}).
Other physically-inspired parametrizations have been proposed. For example, in Refs.~\citenumber{Baldsiefen2013-ob}~and~\citenumber{Lemke2022-vw}, the ONs followed a Fermi–Dirac distribution; however, additional Lagrange multipliers were needed to satisfy Eqs.~(\ref{eqn:lowdin_const}) and (\ref{eq:sum1}). 
It is worth noting that the framework described in Refs.~\citenumber{Baldsiefen2013-ob}~and~\citenumber{Lemke2022-vw} is akin to the ``sigmoid'' activation function.

\begin{eqnarray}
\label{eqn:softmax_grad}
    \dfrac{\partial n_{p}}{\partial \gamma_{r}} &=& \dfrac{\partial \sigma(\gamma_{p})}{\partial \gamma_{r}} \\
    &=&\begin{cases}
        \dfrac{e^{\gamma_p} \left (\sum_q e^{\gamma_q} - e^{\gamma_p}\right )}{\left(\sum_q e^{\gamma_q}\right)^2} & \text{if } r = p;  \nonumber \\ \\[0.1ex]
        \dfrac{-e^{\gamma_p} e^{\gamma_r}}{\left(\sum_q e^{\gamma_q}\right)^2} & \text{otherwise}.
    \end{cases}
\end{eqnarray}

\subsection{Example: A subspace with three orbitals}
To better understand the performance of the ``trigonometric'' and ``softmax'' parameterizations, let us consider a subspace with a single SDONO and two WDONO for three total orbitals. This scenario enables a direct visual comparison of the two parameterization schemes. The trigonometric parameterization for three ONs is,
\begin{eqnarray}
    n_g &=& \tfrac{1}{2}\left(1 + \cos^2 \gamma_g \right) \nonumber\\
    n_{p_1} &=& (1-n_g) \sin^2 \gamma_{p_1} \nonumber\\
    n_{p_2} &=& (1-n_g) \cos^2 \gamma_{p_1}, 
\end{eqnarray}
where only two $\gamma$ auxiliary variables are needed to define the three ONs. Conversely, using the softmax parameterization, the ONs are
\begin{eqnarray}
    n_g &=& \frac{e^{\gamma_{g}}}{e^{\gamma_{g}} + e^{\gamma_{p_1}} + e^{\gamma_{p_2}}} \nonumber \\
    n_{p_1} &=& \frac{e^{\gamma_{p_1}}}{e^{\gamma_{g}} + e^{\gamma_{p_1}} + e^{\gamma_{p_2}}} \nonumber \\
    n_{p_2} &=& \frac{e^{\gamma_{p_2}}}{e^{\gamma_{g}} + e^{\gamma_{p_1}} + e^{\gamma_{p_2}}},
\end{eqnarray}
where three auxiliary variables are needed to determine all three ONs.

\begin{figure}[!htb]
    \centering
    \input{Figures/Surfaces2}
    \caption{The value of ONs in a subspace of three NOs ($n_{g}$, $n_{p_1}$ and $n_{p_2}$) as a function of the auxiliary variables ($\gamma_p$) according to (a) the trigonometric and (b) the softmax parameterizations. The softmax parameterization depends on three $\gamma_p$ variables. We have fixed $\gamma_{p_2}$ to $\frac{1}{2}$ for illustrative purposes.}
    \label{fig:trigonometric-softmax}
\end{figure}
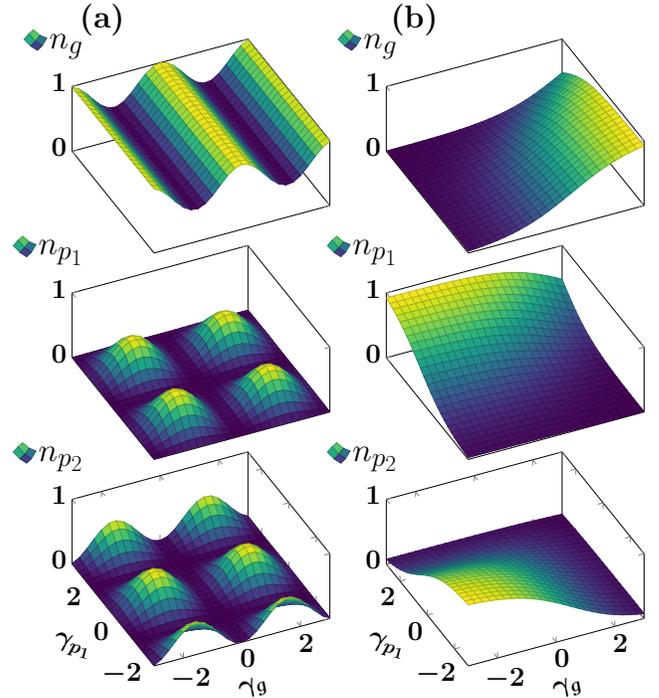

In both approaches, the ONs are parameterized by unconstrained auxiliary variables. However, trigonometric parameterization provides surfaces that feature several valleys and hills, as shown in Fig.~\ref{fig:trigonometric-softmax}(a), where the optimizer can be trapped due to the vanishing of the gradients. Furthermore, if the optimizer takes a step slightly larger than necessary, it may transition to a different basin, invalidating prior steps and destabilizing the optimization process. In contrast, the softmax function provides a better representation of the ONs, as illustrated in Fig.~\ref{fig:trigonometric-softmax}(b), where nonoscillatory surfaces represent the three ONs. In particular, it offers a steady structure for ONs near zero and one, and a nearly linear surface for fractional ONs, which is crucial for optimizing multireference systems.

\section*{Computational Details}
All calculations were performed using the PyNOF package~\cite{pynof:github} with the PNOF7 functional using an extended pairing approach, employing the highest number of $\mathrm{N_g}$ allowed for the def2-TZVPD basis.\cite{Weigend2005-ty,Rappoport2010-cv}. The minimization procedure was carried out with alternate optimizations of NOs and ONs, using a convergence criterion of energy change below $10^{-4}$ Ha for external iterations. Since the PNOF definition assigns a phase to the given strongly occupied orbital, the occupation numbers of each subspace $\Omega_g$ are automatically inspected at the end of each external iteration to ensure that its occupation number has a larger value than the occupations of the weakly occupied orbitals. If the direction of this inequality is not satisfied, the corresponding orbitals and occupation numbers are exchanged, with a negligible computational demand. The optimization of NOs was limited to 30 internal iterations using the orbital rotation method,\cite{Elayan2022-nb} while internal iterations of ONs were performed without any specific limitation on the number of iterations, using a termination criterion of $10^{-5}$ for the norm of the gradient of either the trigonometric or the softmax parameterization. In both cases, the minimization has been carried out using the conjugate gradient method, with derivatives of the energy with respect to $\gamma$ computed by means of the chain rule, that is, multiplying the derivatives of the energy with respect to the occupation numbers with the derivatives of the occupation numbers with respect to $\gamma$. As the purpose of this work is to study specifically the performance of the occupation number parameterization in the minimization of ONs and in the overall calculation, in the following, the optimization cycles for ONs will be termed ``internal iterations", leaving aside the internal iterations of the orbital optimization, which are not relevant to the present work. On the hardware front, simulations were executed on a computer with an Intel Xeon Silver 4208 processor with 16 cores and a GPU NVIDIA GeForce RTX 4090.

\section{Results}
We benchmarked the trigonometric and softmax parameterization methods in the W4-17-MR database, which contains data for 17 multireference chemical systems.\cite{Karton2017-nf} These systems are characterized by significantly fractional occupation numbers, highlighting the complexity of the benchmarking cases. It is crucial to understand that the optimization process involves multiple external iterations. Each of these iterations included optimizing both the orbitals and the occupation numbers, with each step further divided into several internal iterations.

\begin{figure}[htbp]
    \centering
    \input{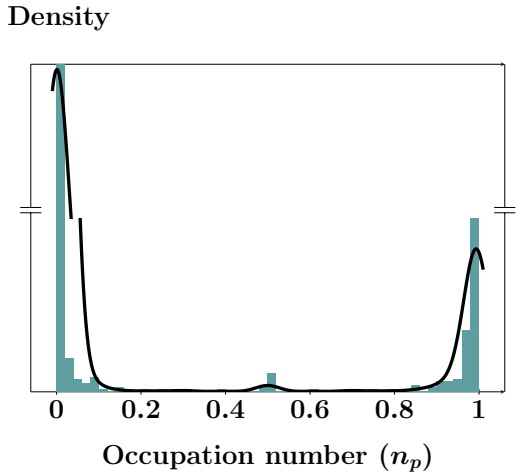}
    \caption{Normalized distribution of the occupation numbers of all molecules in the W4-17-MR dataset as obtained from PNOF7 calculations.}
    \label{fig:distribution_of_ON}
\end{figure}

Figure ~\ref{fig:distribution_of_ON} displays the ONs' distribution across all molecules within the W4-17-MR database, revealing the necessity for specific parametrization across three distinct regimes: $n_{p}\approx 0$, $n_{p}\approx 1$, and $n_{p}\approx \tfrac{1}{2}$. This is particularly important for accurately modeling multireference systems. Both the trigonometric and softmax parameterization methods successfully capture these regimes. However, as evident from Fig.~\ref{fig:trigonometric-softmax}, the softmax function, compared to the trigonometric series, offers a wider space for the auxiliary variables ($\gamma_p$) to represent these conditions accurately.

\begin{figure*}[htbp]
    \centering
    \includegraphics[width=\textwidth]{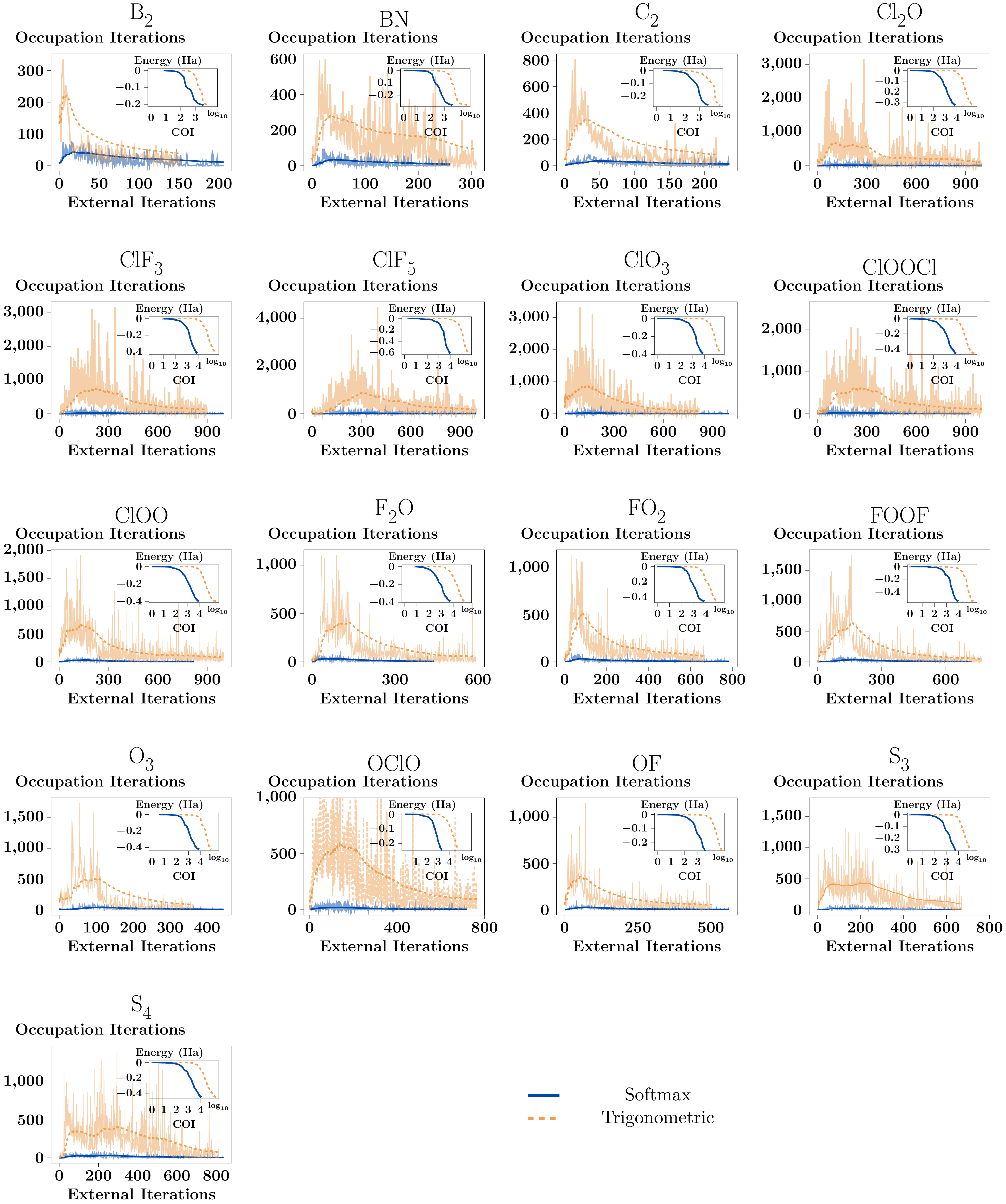}
    \caption{Representative calculation of the W4-17-MR set using the softmax (blue solid curves) and trigonometric (orange dashed curves) parameterization methods. The main panels illustrate the number of occupation iterations for each external iteration. The inset panels depict the energy relative to Hartree-Fock as a function of the cumulative occupation iterations (COI) on a logarithmic scale.}
    \label{fig:w4-17-mr_iterations}
\end{figure*}

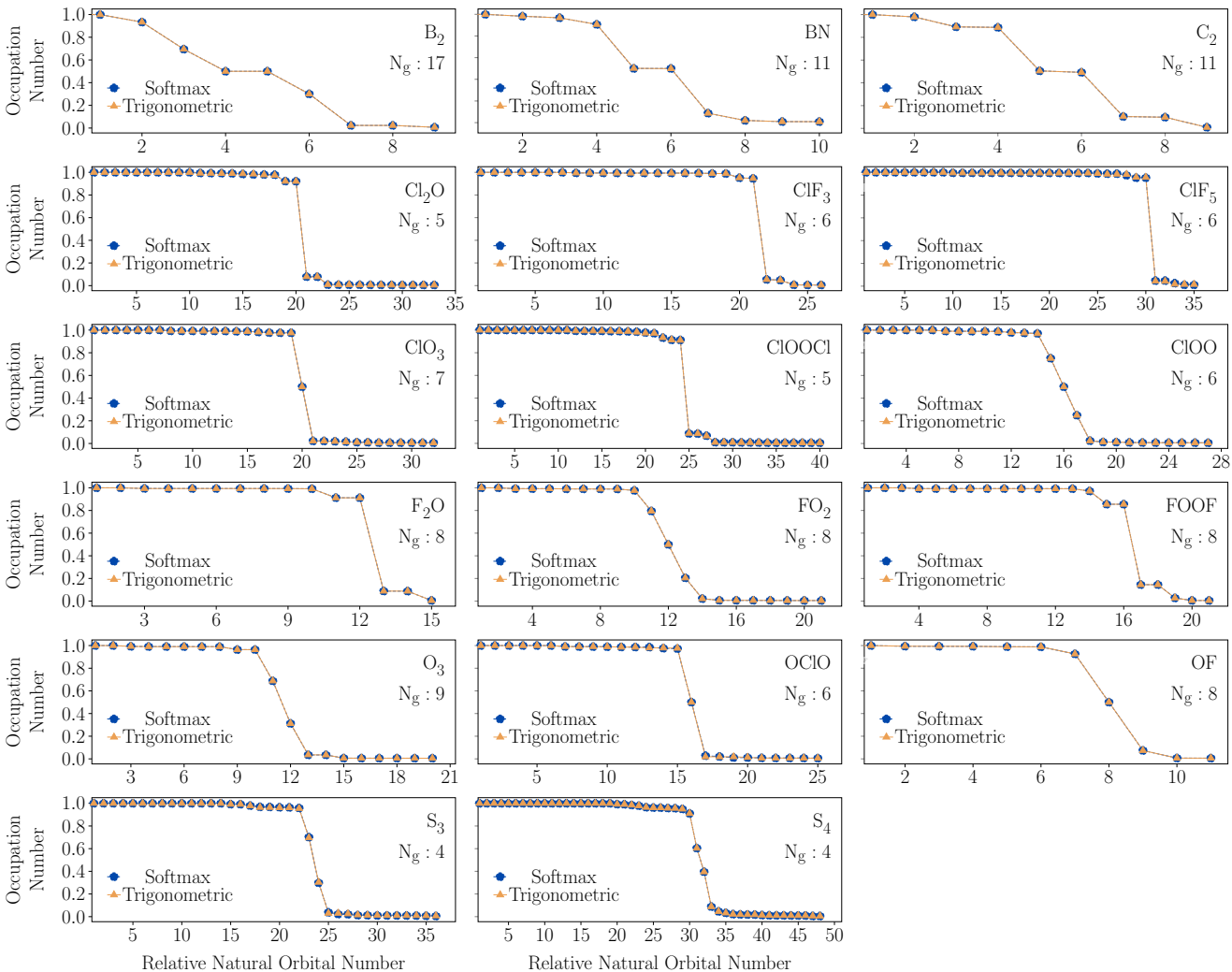
\begin{figure*}[htb]
    \centering
    \input{Figures/Figures_jctc/n_vs_energy_softmax_jctc}
    \caption{The ONs for each molecule in the W4-17-MR set were computed with both the frameworks - softmax (blue) and trigonometric (orange). These figures highlight the propensity of the ONs to cluster near one and zero.}
    \label{fig:n_vs_energy_softmax}
\end{figure*}

In this regard, Fig.~\ref{fig:w4-17-mr_iterations} shows the number of internal iterations required for optimizing the ONs during each external iteration and (inset panel) energy trajectory as a function of cumulative occupation iterations (COI) for all molecules of the W4-17-MR set. As we can observe, softmax parameterization demonstrates superior performance by requiring significantly fewer internal iterations than its trigonometric counterpart. Compared to traditional parameterization methods, the COI panels also reveal that the softmax framework achieves energy convergence with fewer occupation iterations—ranging between one and two orders of magnitude less. Consequently, this indicates that softmax parameterization substantially enhances the overall optimization process.

The performance of the softmax parameterization can be explained by the ONs distribution, as presented in Figs.~\ref{fig:distribution_of_ON} and \ref{fig:n_vs_energy_softmax}, where it can be seen that most ONs are close to one and zero, and only a few have values that are significantly fractional and close to $\tfrac{1}{2}$. This distribution is in good agreement with the concept of core and virtual orbitals, underscoring the existence of ONs firmly entrenched at the extreme ends of the occupancy spectrum, even in a multireference set. In practice, only a few orbitals close to the highest occupied and the lowest unoccupied NO border exhibit deviations in their occupancy, with these fractional occupations becoming critical for strongly correlated systems. The observed distribution supports the adoption of sophisticated parameterization functions, such as the softmax function, from a data-driven perspective.

\begin{figure}[!htb]
    \centering
    \input{Figures/W4-17-MR-dE-M-PNOF7_new}
    \caption{(a) The ratio between the trigonometric and softmax occupancy iterations, $R_\text{T:S}$. (b) The energy difference of occupancy optimization between both parameterizations, $\Delta E = E_{\text{softmax}}-E_{\text{trigonometric}}$, using the same converged set of orbitals for both parameterizations. Both panels are for each molecule of the W4-17-MR set with PNOF7. The color bar shows relative values of the M-diagnostic, which indicates the multireference nature of the systems.}
    \label{fig:mol_vs_delta}
\end{figure}
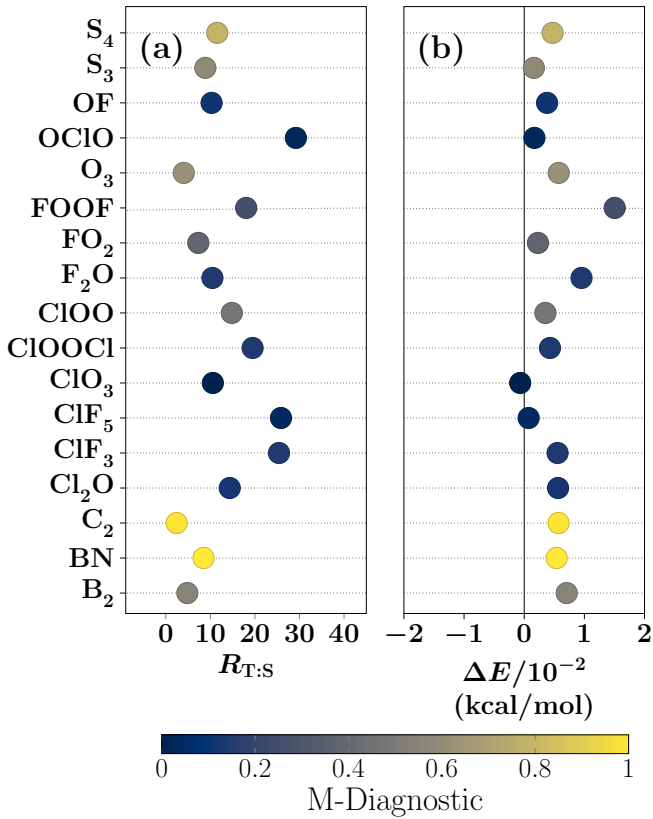

Finally, Fig. \ref{fig:mol_vs_delta} displays the optimization results of ONs, starting from the already converged NOs, with the molecules indexed vertically. The markers' colors correspond to the M-diagnostic developed by Tishchenko \textit{et al}.\cite{Tishchenko2008} and later adapted for the NOF framework \cite{Lew-Yee2022-te}. This color coding illustrates the fractional occupation levels in each system's highest occupied natural orbital. Specifically, Fig. \ref{fig:mol_vs_delta}(a) compares the number of iterations required using the trigonometric and softmax parameterizations ($R_\text{T:S}$), demonstrating a reduction in iterations with the softmax approach across all molecules when the softmax parameterization is used.  
For this purpose, we define the $R_\text{T:S}$ ratio between the number of occupation iterations performed when using the trigonometric and softmax parameterization.
Notably, the $R_\text{T:S}$ ratios are $\sim$ 2, 4, and 5 for \ce{C2}, \ce{O3}, and \ce{B2}, respectively, indicating enhanced convergence with the softmax parameterization, even in these smaller systems. Other molecules show greater acceleration, with the ratio of iterations ranging from 10 to 30, always in favor of the softmax parameterization.
In addition, energy calculations using both frameworks reach comparable values. Fig. \ref{fig:mol_vs_delta}(b) plots the energy differences ($\Delta E = E_\text{softmax} - E_\text{trigonometric}$) along the horizontal axis, where a positive value indicates the trigonometric parameterization reached a lower energy value than the softmax parameterization.
Although the softmax parameterization generally results in slightly higher energy values ($\Delta E>0$), the maximum deviation is minimal, at only 0.0015 kcal/mol. Given the magnitude of these differences, we conclude that the softmax parameterization not only equates in energy to the trigonometric approach but also significantly reduces the number of iterations required, thus enhancing the stability of the optimization process for the ONs.

\section*{Conclusions}
We have introduced an alternative methodology to parameterize the ONs of NOs, a critical component in 1RDMFT methodologies. Our approach leverages the softmax function, a key component in modern deep learning models, to ensure the N-representability constraint without the need for additional variables or complex constraint-optimization strategies.  This new strategy significantly accelerates the convergence of PNOF calculations, achieving a 2 to 30-fold increase in speed compared to traditional parameterizations, such as trigonometric methods.

This method has been validated on the W4-17-MR molecular dataset, known for its multireference character, demonstrating its effectiveness and efficiency. The success of this approach in handling the dataset's complexities highlights its potential applicability across various 1RDMFT methodologies that require optimization of ONs. Moreover, the adoption of alternative parameterizations, such as softmax, could provide groundbreaking solutions to the intricate problems faced in quantum chemistry. This opens new avenues for enhancing computational methodologies and tackling the demanding challenges within the RDMFT field.

\begin{acknowledgement}
R. A. Vargas-Hernández acknowledges useful discussions with Chong Sun.
M. Piris thanks the Eusko Jaurlaritza (Basque Government), Ref. No. : IT1584-22, for financial support, as well as Grant No. PID 2021-126714NB-I00, funded by Grant No. MCIN/AEI/10.13039/501100011033. J. F. H. Lew-Yee acknowledges the Donostia International Physics Center (DIPC) and the MCIN program ``Severo Ochoa'' under reference No. AEI/ CEX2018-000867-S for post-doctoral funding (Ref. No. : 2023/74). L. Franco with CVU Grant No. 1312732  acknowledges ``Consejo Nacional de Humanidades, Ciencias y Tecnologías (CONAHCyT)'' for a master scholarship. J. M. del Campo, L. Franco, and I. A. Bonfil-Rivera acknowledge funding with project Grant No. IN201822 from PAPIIT, and computing resources from ``Laboratorio Nacional de Cómputo de Alto Desempeño (LANCAD)'' with project Grant No. LANCAD-UNAMDGTIC-270. 
\end{acknowledgement}

\bibliography{bibliography}

\end{document}

%% file: Figures/pairing-scheme-F.tex

\definecolor{color0}{HTML}{891e35}
\definecolor{color1}{HTML}{001a57}
\definecolor{color2}{HTML}{9c63b0}
\definecolor{color3}{HTML}{3b7861}
\definecolor{color4}{HTML}{5b3a29}


\begin{tikzpicture}[scale=0.6]



\newcommand{\len}{3.5pt}
\newcommand{\w}{6pt}

\node[color0] at (2.5,0.8) {$\Omega_1$};
\draw[->,color0, very thick,rounded corners] (2.0,0.4) -- (5.7,0.4) -- (5.7,11.75) -- (2.0,11.75);
\draw[color0, thick, dashed,fill=color0!20] (-0.2,11.25) rectangle (1.7,12.65);

\draw[arrows = {-Stealth[harpoon,length=\len,width=\w]}, very thick] (1.0,12.35) -- (1.0,12.15); 
\draw[arrows = {-Stealth[harpoon,length=\len,width=\w]}, very thick] (0.5,12.15) -- (0.5,12.35); 
\draw[color0, very thick] (0.0,12.05) -- (1.5,12.05);

\draw[arrows = {-Stealth[harpoon,length=\len,width=\w]}, very thick] (1.0,11.75) -- (1.0,11.55); 
\draw[arrows = {-Stealth[harpoon,length=\len,width=\w]}, very thick] (0.5,11.55) -- (0.5,11.75); 
\draw[color0, very thick] (0.0,11.45) -- (1.5,11.45);


\node[color1] at (2.5,2.0) {$\Omega_2$};
\draw[->,color1, very thick,rounded corners] (2.0,1.6) -- (4.87,1.6) -- (4.87,10.15) -- (2.0,10.15);
\draw[color1, thick, dashed,fill=color1!20] (-0.2,9.65) rectangle (1.7,11.05);

\draw[arrows = {-Stealth[harpoon,length=\len,width=\w]}, very thick] (1.0,10.75) -- (1.0,10.55);
\draw[arrows = {-Stealth[harpoon,length=\len,width=\w]}, very thick] (0.5,10.55) -- (0.5,10.75);
\draw[color1, very thick] (0.0,10.45) -- (1.5,10.45);

\draw[arrows = {-Stealth[harpoon,length=\len,width=\w]}, very thick] (1.0,10.15) -- (1.0,9.95);
\draw[arrows = {-Stealth[harpoon,length=\len,width=\w]}, very thick] (0.5,9.95) -- (0.5,10.15);
\draw[color1, very thick] (0.0,9.85) -- (1.5,9.85);


\node[color2] at (2.5,3.2) {$\Omega_3$};
\draw[->,color2, very thick, rounded corners] (2.0,2.8) -- (4.03,2.8) -- (4.03,8.55) -- (2.0,8.55);
\draw[color2, thick,dashed,fill=color2!20] (-0.2,8.05) rectangle (1.7, 9.45);

\draw[arrows = {-Stealth[harpoon,length=\len,width=\w]}, very thick] (1.0,9.15) -- (1.0,8.95);
\draw[arrows = {-Stealth[harpoon,length=\len,width=\w]}, very thick] (0.5,8.95) -- (0.5,9.15);
\draw[color2, very thick] (0.0,8.85) -- (1.5,8.85);

\draw[arrows = {-Stealth[harpoon,length=\len,width=\w]}, very thick] (1.0,8.55) -- (1.0,8.35);
\draw[arrows = {-Stealth[harpoon,length=\len,width=\w]}, very thick] (0.5,8.35) -- (0.5,8.55);
\draw[color2, very thick] (0.0,8.25) -- (1.5,8.25);


\node[color3] at (2.5,4.4) {$\Omega_4$};
\draw[->,color3, very thick,rounded corners] (2.0,4.0) -- (3.2,4.0) -- (3.2,6.95) -- (2.0,6.95);
\draw[color3, thick,dashed,fill=color3!20] (-0.2,6.45) rectangle (1.7,7.85); 

\draw[arrows = {-Stealth[harpoon,length=\len,width=\w]}, very thick] (1.0,7.64) -- (1.0,7.35);
\draw[arrows = {-Stealth[harpoon,length=\len,width=\w]}, very thick] (0.5,7.35) -- (0.5,7.64);
\draw[color3, very thick] (0.0,7.25) -- (1.5,7.25); %

\draw[arrows = {-Stealth[harpoon,length=\len,width=\w]}, very thick] (1.0,7.03) -- (1.0,6.75);
\draw[arrows = {-Stealth[harpoon,length=\len,width=\w]}, very thick] (0.5,6.75) -- (0.5,7.03);
\draw[color3, very thick] (0.0,6.65) -- (1.5,6.65); %


\node[black] at (-1.2,5.4) {\scriptsize $\mathrm{N_{\Omega}}$};
\draw[black, thick] (-0.4,6.25) -- (1.9,6.25);

\draw[color4, very thick, dotted,fill=color4!20] (-0.2,5.2) rectangle (1.7,6.05);

\draw[arrows = {-Stealth[harpoon,length=4.5pt,width=5pt]}, very thick] (1.0,5.85) -- (1.0,5.5);
\draw[arrows = {-Stealth[harpoon,length=4.5pt,width=5pt]}, very thick] (0.5,5.5) -- (0.5,5.85);
\draw[color4, very thick] (0.0,5.4) -- (1.5,5.4);

\node[color4] at (2.5,5.6) {$\Omega_5$};

\node[black] at (-1.2,4.1) {\scriptsize $\mathrm{N_{II}/2}$};
\draw[black, thick] (-0.4,5.0) -- (1.9,5.0);

            
\draw[color3,very thick,fill=color3!20] (-0.2,3.8) rectangle (1.7,4.8); 
\draw[arrows = {-Stealth[harpoon,length=5.7pt,width=5pt]}, very thick] (1.0,4.6-0.06) -- (1.0,4.1);
\draw[arrows = {-Stealth[harpoon,length=5.7pt,width=5pt]}, very thick] (0.5,4.1) -- (0.5,4.6-0.06); 
\draw[color3, very thick] (0.0,4.0) -- (1.5,4.0);

\draw[color2,very thick,fill=color2!20] (-0.2,2.6) rectangle (1.7,3.6);
\draw[arrows = {-Stealth[harpoon,length=5.7pt,width=5pt]}, very thick] (1.0,3.4) -- (1.0,2.9);
\draw[arrows = {-Stealth[harpoon,length=5.7pt,width=5pt]}, very thick] (0.5,2.9) -- (0.5,3.4);
\draw[color2, very thick] (0.0,2.8) -- (1.5,2.8);

\draw[color1,very thick,fill=color1!20] (-0.2,1.4) rectangle (1.7,2.4);
\draw[arrows = {-Stealth[harpoon,length=5.7pt,width=5pt]}, very thick] (1.0,2.2) -- (1.0,1.7);
\draw[arrows = {-Stealth[harpoon,length=5.7pt,width=5pt]}, very thick] (0.5,1.7) -- (0.5,2.2);
\draw[color1, very thick] (0.0,1.6) -- (1.5,1.6);

\draw[color0,very thick,fill=color0!20] (-0.2,0.2) rectangle (1.7,1.2);
\draw[arrows = {-Stealth[harpoon,length=5.7pt,width=5pt]}, very thick] (1.0,1.0) -- (1.0,0.5);
\draw[arrows = {-Stealth[harpoon,length=5.7pt,width=5pt]}, very thick] (0.5,0.5) -- (0.5,1.0);
\draw[color0, very thick] (0.0,0.4) -- (1.5,0.4);

\node at (-6.25,9.55) {\footnotesize WDONO};
\node at (-6.25,8.8)  {\footnotesize $0 \leq n_p \leq 1/2$};
\node at (-3.25,9.55) {\footnotesize $\Omega_{\mathrm{II}}$};
\draw [decorate,decoration={brace,amplitude=10pt},xshift=-4pt,yshift=0pt] (-1.8,6.45) -- (-1.8,12.65) node [black,midway,xshift=-0.6cm] {};

\node at (-6.25,5.8) {\footnotesize SONO};
\node at (-6.25,5.05) {\footnotesize $n_p = 1/2$};
\node at (-3.25,5.8) {\footnotesize $\Omega_{\mathrm{I}}$};
\draw [decorate,decoration={brace,amplitude=10pt},xshift=-4pt,yshift=0pt] (-1.8,5.2) -- (-1.8,6.25) node [black,midway,xshift=-0.6cm] {};

\node at (-6.25,2.5) {\footnotesize SDONO};
\node at (-6.25,1.75) {\footnotesize $1/2 \leq n_p \leq 1$};
\node at (-3.25,2.5) {\footnotesize $\Omega_{\mathrm{II}}$};
\draw [decorate,decoration={brace,amplitude=10pt},xshift=-4pt,yshift=0pt] (-1.8,0.2) -- (-1.8,4.8) node [black,midway,xshift=-0.6cm] {};

\newcommand{\x}{-1.65}
\newcommand{\y}{-1.5}

\fill[blue!10!white] (\x-1.1,\y+0.45) rectangle (\x+7.4,\y+1.3);
\draw[thick,rounded corners] (\x-1.1,\y-0.5) rectangle (\x+7.4,\y+1.3);
\draw[dotted, thick] (\x-1.1,\y+0.45) -- (\x+7.4,\y+0.45);

\node at (\x,\y)     {\scriptsize $2\sum_p n_p$};
\node at (\x+1.5,\y) {\scriptsize $2$};
\node at (\x+2.5,\y) {\scriptsize $2$};
\node at (\x+3.5,\y) {\scriptsize $2$};
\node at (\x+4.5,\y) {\scriptsize $2$};
\node at (\x+5.5,\y) {\scriptsize $1$};

\node at         (\x,    \y+0.8) {\scriptsize Subspace}  ;
\node[color0] at (\x+1.5,\y+0.8) {\scriptsize $\Omega_1$};
\node[color1] at (\x+2.5,\y+0.8) {\scriptsize $\Omega_2$};
\node[color2] at (\x+3.5,\y+0.8) {\scriptsize $\Omega_3$};
\node[color3] at (\x+4.5,\y+0.8) {\scriptsize $\Omega_4$};
\node[color4] at (\x+5.5,\y+0.8) {\scriptsize $\Omega_5$};

\node at (\x+6.5,\y+0.8) {\scriptsize Total};
\node at (\x+6.5,\y)     {\scriptsize $9$}; 

\end{tikzpicture}   

%% file: Figures/Surfaces2.tex
\begin{tikzpicture}[scale=0.5]

\begin{groupplot}[group style={group name=surfaces, group size=2 by 3,vertical sep=-0.2cm, horizontal sep=1.5cm}, view={-25}{60}]

\nextgroupplot[
            legend style={font=\huge, inner sep=9pt, at={(0.1,1.1)}, draw=none, fill=none},
            domain=-3:3,
            domain y=-3:3,
            zmin=0, zmax=1,
            xlabel style={xshift=-5pt, yshift=5pt, rotate=25},
            xtick=\empty,
            ytick=\empty,
            ztick={0,1},
            ylabel style={xshift=20pt, yshift=12pt, rotate=-15},
            label style={font=\huge},
            tick label style={font=\LARGE},
            xlabel style={align=center, inner sep=0pt, font=\LARGE\bfseries\boldmath},
            ylabel style={align=center, inner sep=5pt, font=\LARGE\bfseries\boldmath},
            zlabel style={align=center, inner sep=5pt, font=\LARGE\bfseries\boldmath},
            x tick label style={font=\LARGE\bfseries\boldmath},
            y tick label style={font=\LARGE\bfseries\boldmath},
            z tick label style={font=\LARGE\bfseries\boldmath},
            colormap/viridis,
        ]
        \addplot3[
            surf,
        ] 
        {1/2*(1+cos(deg(x))^2)};
        \addlegendentry{$n_g$} 

\nextgroupplot[
            legend style={font=\huge, inner sep=9pt, at={(0.1,1.1)}, draw=none, fill=none},
            domain=-3:3,
            domain y=-3:3,
            zmin=0, zmax=1,
            xlabel style={xshift=-5pt, yshift=5pt, rotate=25},
            xtick=\empty,
            ytick=\empty,
            ztick={0,1},
            ylabel style={xshift=20pt, yshift=12pt, rotate=-15},
            label style={font=\huge},
            tick label style={font=\LARGE},
            xlabel style={align=center, inner sep=0pt, font=\LARGE\bfseries\boldmath},
            ylabel style={align=center, inner sep=5pt, font=\LARGE\bfseries\boldmath},
            zlabel style={align=center, inner sep=5pt, font=\LARGE\bfseries\boldmath},
            x tick label style={font=\LARGE\bfseries\boldmath},
            y tick label style={font=\LARGE\bfseries\boldmath},
            z tick label style={font=\LARGE\bfseries\boldmath},
            colormap/viridis,
        ]
        \addplot3[
            surf,
        ] 
        {exp(x)/(exp(x)+exp(y)+exp(1/2))};
        \addlegendentry{$n_g$}

\nextgroupplot[
            legend style={font=\huge, inner sep=9pt, at={(0.1,1.1)}, draw=none, fill=none},
            domain=-3:3,
            domain y=-3:3,
            zmin=0, zmax=1,
            xlabel style={xshift=-5pt, yshift=5pt, rotate=25},
            xtick=\empty,
            ytick=\empty,
            ztick={0,1},
            ylabel style={xshift=20pt, yshift=12pt, rotate=-15},
            label style={font=\huge},
            tick label style={font=\LARGE},
            xlabel style={align=center, inner sep=0pt, font=\LARGE\bfseries\boldmath},
            ylabel style={align=center, inner sep=5pt, font=\LARGE\bfseries\boldmath},
            zlabel style={align=center, inner sep=5pt, font=\LARGE\bfseries\boldmath},
            x tick label style={font=\LARGE\bfseries\boldmath},
            y tick label style={font=\LARGE\bfseries\boldmath},
            z tick label style={font=\LARGE\bfseries\boldmath},
            colormap/viridis,
        ]
        \addplot3[
            surf,
        ] 
        {(1-1/2*(1+cos(deg(x))^2))*sin(deg(y))^2};
        \addlegendentry{$n_{p_{1}}$}

\nextgroupplot[
            legend style={font=\huge, inner sep=9pt, at={(0.1,1.1)}, draw=none, fill=none},
            domain=-3:3,
            domain y=-3:3,
            zmin=0, zmax=1,
            xlabel style={xshift=-5pt, yshift=5pt, rotate=25},
            xtick=\empty,
            ytick=\empty,
            ztick={0,1},
            ylabel style={xshift=20pt, yshift=12pt, rotate=-15},
            label style={font=\huge},
            tick label style={font=\LARGE},
            xlabel style={align=center, inner sep=0pt, font=\LARGE\bfseries\boldmath},
            ylabel style={align=center, inner sep=5pt, font=\LARGE\bfseries\boldmath},
            zlabel style={align=center, inner sep=5pt, font=\LARGE\bfseries\boldmath},
            x tick label style={font=\LARGE\bfseries\boldmath},
            y tick label style={font=\LARGE\bfseries\boldmath},
            z tick label style={font=\LARGE\bfseries\boldmath},
            colormap/viridis,
        ]
        \addplot3[
            surf,
        ] 
        {exp(y)/(exp(x)+exp(y)+exp(1/2))};
        \addlegendentry{$n_{p_{1}}$}

\nextgroupplot[
            legend style={font=\huge, inner sep=9pt, at={(0.1,1.1)}, draw=none, fill=none},
            domain=-3:3,
            domain y=-3:3,
            zmin=0, zmax=1,
            xlabel = {$\gamma_{g}$},
            xlabel style={xshift=-5pt, yshift=5pt, rotate=25},
            xtick={-2,0,2}, 
            ztick={0,1},
            ylabel = {$\gamma_{p_{1}}$},
            ylabel style={xshift=10pt, yshift=30pt, rotate=-15},
            label style={font=\huge},
            tick label style={font=\LARGE},
            xlabel style={align=center, inner sep=0pt, font=\LARGE\bfseries\boldmath},
            ylabel style={align=center, inner sep=5pt, font=\LARGE\bfseries\boldmath},
            zlabel style={align=center, inner sep=5pt, font=\LARGE\bfseries\boldmath},
            x tick label style={font=\LARGE\bfseries\boldmath},
            y tick label style={font=\LARGE\bfseries\boldmath},
            z tick label style={font=\LARGE\bfseries\boldmath},
            colormap/viridis,
        ]
        \addplot3[
            surf,
        ] 
        {(1-1/2*(1+cos(deg(x))^2))*cos(deg(y))^2};
        \addlegendentry{$n_{p_{2}}$}

\nextgroupplot[
            legend style={font=\huge, inner sep=9pt, at={(0.1,1.1)}, draw=none, fill=none},
            domain=-3:3,
            domain y=-3:3,
            zmin=0, zmax=1,
            xlabel = {$\gamma_{g}$},
            xlabel style={xshift=-5pt, yshift=5pt, rotate=25},
            xtick={-2,0,2}, 
            ztick={0,1}, 
            ylabel = {$\gamma_{p_{1}}$},
            ylabel style={xshift=10pt, yshift=30pt, rotate=-15},
            label style={font=\huge},
            tick label style={font=\LARGE},
            xlabel style={align=center, inner sep=0pt, font=\LARGE\bfseries\boldmath},
            ylabel style={align=center, inner sep=5pt, font=\LARGE\bfseries\boldmath},
            zlabel style={align=center, inner sep=5pt, font=\LARGE\bfseries\boldmath},
            x tick label style={font=\LARGE\bfseries\boldmath},
            y tick label style={font=\LARGE\bfseries\boldmath},
            z tick label style={font=\LARGE\bfseries\boldmath},
            colormap/viridis,
        ]
        \addplot3[
            surf,
        ] 
        {exp(1/2)/(exp(x)+exp(y)+exp(1/2))};
        \addlegendentry{$n_{p_{2}}$}

\end{groupplot}

\node[align=center, font=\normalsize] at (0.8,6.2) {\bf{(a)}};
\node[align=center, font=\normalsize] at (9.1,6.2) {\bf{(b)}};

\end{tikzpicture}

%% file: Figures/Figures_jctc/n_vs_energy_softmax_jctc.tex


\pgfplotsset{
    every non boxed x axis/.style={} 
}

\begin{tikzpicture}[scale = 0.32]

\definecolor{brown}{RGB}{236,160,83}
\definecolor{blue}{RGB}{0,71,171}
\definecolor{cadetblue}{RGB}{95,158,160}

\begin{groupplot}[group style={group size=3 by 7, horizontal sep=1cm, vertical sep=1.5cm}, y tick label style={/pgf/number format/.cd,fixed,fixed zerofill,precision=1,/tikz/.cd},]

\nextgroupplot[
tick align=outside,
tick pos=left,
x grid style={darkgray176},
xmin=0.8, xmax=9.5,
xtick style={color=black},
xticklabel style = {font=\huge},
xtick distance=2,
y grid style={darkgray176},
ylabel={Occupation \\ Number},
ylabel style={align=center, yshift=15pt, font=\huge},
ymin=-0.043078425, ymax=1.049233125,
ytick style={color=black},
yticklabel style = {font=\huge},
width=\textwidth,
mark size=5pt,
height=0.4\textwidth,
legend style={draw=none, at={(0.4,0.45)},fill=none, font=\Huge},
]

\addplot+[draw=blue, dash pattern=on 4pt off 2pt, legend image post style={sharp plot, dash pattern=on 4pt off 2pt}, mark options={fill=blue}]
table{%
x  y
0	0.99953985
1	0.9994919
2	0.9331074
3	0.69471435
4	0.5
5	0.5
6	0.3007961
7	0.02310105
8	0.023097
9	0.0068179
};
\addplot+[draw=brown, mark=triangle*, mark options={fill=brown}, legend image post style={sharp plot}]
table{%
x  y
0	0.9995398
1	0.99948955
2	0.93307065
3	0.6948684
4	0.5
5	0.5
6	0.3006782
7	0.0231615
8	0.0231405
9	0.0068591
};
\legend{Softmax, Trigonometric}
\node[draw=none, anchor=north east, font=\Huge] at (rel axis cs:0.98, 0.9) {\ce{B2}};
\node[draw=none, anchor=north east, font=\Huge] at (rel axis cs:0.98, 0.65) {$\mathrm{N_g}:17$};

\nextgroupplot[
tick align=outside,
tick pos=left,
x grid style={darkgray176},
xmin=0.8, xmax=10.6,
xtick style={color=black},
xticklabel style = {font=\huge},
yticklabel style = {font=\huge},
xtick distance=2,
yticklabels={}
ymin=-0.0405632275, ymax=1.0493877775,
mark size=5pt,
width=\textwidth,
height=0.4\textwidth,
legend style={draw=none, at={(0.4,0.45)},fill=none, font=\Huge},
]

\addplot+[draw=blue, dash pattern=on 4pt off 2pt, legend image post style={sharp plot, dash pattern=on 4pt off 2pt}, mark options={fill=blue}]
table{%
0	0.9998445
1	0.99955605
2	0.98221825
3	0.96766805
4	0.90653135
5	0.5
6	0.49986105
7	0.08716105
8	0.02011985
9	0.00927875
10	0.0089747
};
\addplot+[draw=brown, mark=triangle*, mark options={fill=brown}, legend image post style={sharp plot}]
table {%
0	0.99984445
1	0.9995545
2	0.9822189
3	0.96763195
4	0.90659125
5	0.50148945
6	0.4983955
7	0.0871343
8	0.02016565
9	0.00926445
10	0.0089671
};
\legend{Softmax, Trigonometric}
\node[draw=none, anchor=north east, font=\Huge] at (rel axis cs:0.98, 0.9) {\ce{BN}};
\node[draw=none, anchor=north east, font=\Huge] at (rel axis cs:0.98, 0.65) {$\mathrm{N_g}:11$};

\nextgroupplot[
tick align=outside,
tick pos=left,
x grid style={darkgray176},
xmin=0.8, xmax=9.5,
xticklabel style = {font=\huge},
yticklabel style = {font=\huge},
xtick distance=2,
yticklabels={}
xtick style={color=black},
ymin=-0.04010676, ymax=1.04927736,
mark size=5pt,
mark size=5pt,
width=\textwidth,
height=0.4\textwidth,
legend style={draw=none, at={(0.4,0.45)},fill=none, font=\Huge},
]

\addplot+[draw=blue, dash pattern=on 4pt off 2pt, legend image post style={sharp plot, dash pattern=on 4pt off 2pt}, mark options={fill=blue}]
table{%
x  y
0	0.9997608
1	0.9997504
2	0.979269
3	0.8917862
4	0.8873695
5	0.505101
6	0.49216755
7	0.10420125
8	0.09807785
9	0.0094201
};
\addplot+[draw=brown, mark=triangle*, mark options={fill=brown}, legend image post style={sharp plot}]
table {%
0	0.9997597
1	0.99975095
2	0.97922655
3	0.89170735
4	0.88727725
5	0.50510145
6	0.49219865
7	0.1042554
8	0.0981249
9	0.00941685
};
\legend{Softmax, Trigonometric}
\node[draw=none, anchor=north east, font=\Huge] at (rel axis cs:0.98, 0.9) {\ce{C2}};
\node[draw=none, anchor=north east, font=\Huge] at (rel axis cs:0.98, 0.65) {$\mathrm{N_g}:11$};

\nextgroupplot[
tick align=outside,
tick pos=left,
x grid style={darkgray176},
xmin=0.8, xmax=35,
xticklabel style = {font=\huge},
yticklabel style = {font=\huge},
xtick distance=5,
xtick style={color=black},
y grid style={darkgray176},
ylabel={Occupation \\ Number},
ylabel style={align=center, yshift=15pt, font=\huge},
ymin=-0.0444822475, ymax=1.0497371975,
ytick style={color=black},
mark size=5pt,
width=\textwidth,
height=0.4\textwidth,
legend style={draw=none, at={(0.4,0.45)},fill=none, font=\Huge},
]
\addplot+[draw=blue, dash pattern=on 4pt off 2pt, legend image post style={sharp plot, dash pattern=on 4pt off 2pt}, mark options={fill=blue}]
table{%
x  y
0	0.99999995
1	0.9999999
2	0.99994915
3	0.99992785
4	0.99989885
5	0.9998741
6	0.99987325
7	0.999858
8	0.9998237
9	0.99976035
10	0.9997551
11	0.9937396
12	0.9925465
13	0.99203985
14	0.9905896
15	0.98488485
16	0.981094
17	0.98042945
18	0.978095
19	0.92208505
20	0.92079785
21	0.0788295
22	0.0776272
23	0.0082759
24	0.00784985
25	0.0074479
26	0.00734885
27	0.0064896
28	0.00645945
29	0.0058232
30	0.00570775
31	0.0055585
32	0.00545455
33	0.0053928
};
\addplot+[draw=brown, mark=triangle*, mark options={fill=brown}, legend image post style={sharp plot}]
table{%
x  y
0	0.99999995
1	0.9999999
2	0.99991015
3	0.99990265
4	0.9998989
5	0.99989105
6	0.99988775
7	0.99987695
8	0.9998726
9	0.9997538
10	0.99974605
11	0.9928229
12	0.99176635
13	0.99033245
14	0.9900535
15	0.98760465
16	0.98519275
17	0.97805475
18	0.97466715
19	0.9210086
20	0.92041705
21	0.0792737
22	0.078671
23	0.00941565
24	0.0087533
25	0.00840835
26	0.0070583
27	0.0069462
28	0.0064879
29	0.0057953
30	0.0054858
31	0.0054489
32	0.0054455
33	0.0053156
};
\legend{Softmax, Trigonometric}
\node[draw=none, anchor=north east, font=\Huge] at (rel axis cs:0.98, 0.9) {\ce{Cl2O}};
\node[draw=none, anchor=north east, font=\Huge] at (rel axis cs:0.98, 0.65) {$\mathrm{N_g}:5$};

\nextgroupplot[
tick align=outside,
tick pos=left,
x grid style={darkgray176},
xmin=0.8, xmax=27.5,
xticklabel style = {font=\huge},
yticklabel style = {font=\huge},
xtick distance=5,
yticklabels={},
xtick style={color=black},
ymin=-0.0447363475, ymax=1.0497492975,
mark size=5pt,
width=\textwidth,
height=0.4\textwidth,
legend style={draw=none, at={(0.4,0.45)},fill=none, font=\Huge},
]

\addplot+[draw=blue, dash pattern=on 4pt off 2pt, legend image post style={sharp plot, dash pattern=on 4pt off 2pt}, mark options={fill=blue}]
table{%
x  y
0	0.99999995
1	0.99993965
2	0.9999312
3	0.99993
4	0.9999284
5	0.99990215
6	0.9998014
7	0.9997717
8	0.9945349
9	0.9944689
10	0.9940621
11	0.9936549
12	0.9935964
13	0.9935738
14	0.9934963
15	0.99343605
16	0.99341925
17	0.99276975
18	0.98943745
19	0.98905855
20	0.9515554
21	0.9458094
22	0.0541004
23	0.04836145
24	0.00680265
25	0.0051773
26	0.00515395
};
\addplot+[draw=brown, mark=triangle*, mark options={fill=brown}, legend image post style={sharp plot}]
table {%
0	0.99999995
1	0.99994385
2	0.9999311
3	0.99992995
4	0.99992865
5	0.999889
6	0.99980175
7	0.9997777
8	0.9946353
9	0.99443755
10	0.99401905
11	0.9936899
12	0.99359465
13	0.99354705
14	0.99351915
15	0.99348005
16	0.99343465
17	0.992639
18	0.9893827
19	0.9891408
20	0.95177195
21	0.94619395
22	0.0537227
23	0.04814535
24	0.00693995
25	0.00513465
26	0.0051264
};
\legend{Softmax, Trigonometric}
\node[draw=none, anchor=north east, font=\Huge] at (rel axis cs:0.98, 0.9) {\ce{ClF3}};
\node[draw=none, anchor=north east, font=\Huge] at (rel axis cs:0.98, 0.65) {$\mathrm{N_g}:6$};

\nextgroupplot[
tick align=outside,
tick pos=left,
x grid style={darkgray176},
xmin=0.8, xmax=38.5,
xticklabel style = {font=\huge},
yticklabel style = {font=\huge},
xtick distance=5,
yticklabels={},
xtick style={color=black},
ymin=-0.039775305, ymax=1.049513005,
mark size=5pt,
width=\textwidth,
height=0.4\textwidth,
legend style={draw=none, at={(0.4,0.45)},fill=none, font=\Huge},
]

\addplot+[draw=blue, dash pattern=on 4pt off 2pt, legend image post style={sharp plot, dash pattern=on 4pt off 2pt}, mark options={fill=blue}]
table{%
x  y
0	0.9999999
1	0.9999348
2	0.99993105
3	0.999931
4	0.99993055
5	0.9999291
6	0.99992905
7	0.9998231
8	0.99982155
9	0.99980915
10	0.99520735
11	0.9948071
12	0.99469895
13	0.99469035
14	0.9946601
15	0.99444805
16	0.99435475
17	0.9943217
18	0.9942732
19	0.9942697
20	0.99420265
21	0.993992
22	0.99391795
23	0.9937305
24	0.99366575
25	0.99032075
26	0.9876738
27	0.98752195
28	0.9741396
29	0.95205195
30	0.9520101
31	0.0478819
32	0.0478455
33	0.02567165
34	0.01147235
35	0.01140915
};
\addplot+[draw=brown, mark=triangle*, mark options={fill=brown}, legend image post style={sharp plot}]
table {%
0	0.9999999
1	0.99994875
2	0.99993125
3	0.9999306
4	0.99993055
5	0.9999294
6	0.99992935
7	0.99982645
8	0.9998018
9	0.99979645
10	0.99534815
11	0.99484005
12	0.9947343
13	0.9946997
14	0.99459435
15	0.99445635
16	0.9944396
17	0.9943186
18	0.99429095
19	0.9942287
20	0.9940753
21	0.9939444
22	0.99393795
23	0.9938847
24	0.9937236
25	0.9901973
26	0.9843426
27	0.98417475
28	0.98200985
29	0.95694305
30	0.9557223
31	0.0441761
32	0.0428594
33	0.017818
34	0.01553565
35	0.01527865
};
\legend{Softmax, Trigonometric}
\node[draw=none, anchor=north east, font=\Huge] at (rel axis cs:0.98, 0.9) {\ce{ClF5}};
\node[draw=none, anchor=north east, font=\Huge] at (rel axis cs:0.98, 0.65) {$\mathrm{N_g}:6$};

\draw (axis cs:0.95,0.9) node[
  scale=0.6,
  fill=white,
  draw=none,
  line width=0.4pt,
  inner sep=3.3pt,
  fill opacity=0.5,
  anchor=north east,
  text=black,
  rotate=0.0
]{$ClF_5$};

\nextgroupplot[
tick align=outside,
tick pos=left,
x grid style={darkgray176},
xmin=0.8, xmax=34,
xticklabel style = {font=\huge},
yticklabel style = {font=\huge},
xtick distance=5,
xtick style={color=black},
y grid style={darkgray176},
ylabel={Occupation \\ Number},
ylabel style={align=center, yshift=15pt, font=\huge},
ymin=-0.04354843, ymax=1.04969273,
ytick style={color=black},
mark size=5pt,
width=\textwidth,
height=0.4\textwidth,
legend style={draw=none, at={(0.4,0.45)},fill=none, font=\Huge},
]

\addplot+[draw=blue, dash pattern=on 4pt off 2pt, legend image post style={sharp plot, dash pattern=on 4pt off 2pt}, mark options={fill=blue}]
table{%
x  y
0	0.99999995
1	0.9999018
2	0.99990165
3	0.9999015
4	0.9998524
5	0.99985185
6	0.9998453
7	0.9998415
8	0.9925087
9	0.99178255
10	0.99172195
11	0.991492
12	0.99141565
13	0.99093925
14	0.9885988
15	0.98816975
16	0.98230415
17	0.977338
18	0.975759
19	0.97506115
20	0.5
21	0.0223633
22	0.0218997
23	0.0185516
24	0.01569095
25	0.0104055
26	0.00998535
27	0.00683765
28	0.00682855
29	0.0067494
30	0.00652425
31	0.00636525
32	0.00600555
};
\addplot+[draw=brown, mark=triangle*, mark options={fill=brown}, legend image post style={sharp plot}]
table {%
0	0.99999995
1	0.99990185
2	0.9999016
3	0.99990145
4	0.99985215
5	0.99985185
6	0.99984545
7	0.99984225
8	0.99244835
9	0.99181015
10	0.9917382
11	0.9914264
12	0.9914186
13	0.9909154
14	0.98862
15	0.9880007
16	0.98312615
17	0.9773204
18	0.9755894
19	0.9750265
20	0.5
21	0.0225284
22	0.0219165
23	0.0185569
24	0.01488555
25	0.0105864
26	0.0100262
27	0.0068423
28	0.0068302
29	0.0067395
30	0.00650515
31	0.00640505
32	0.00604325
};
\legend{Softmax, Trigonometric}
\node[draw=none, anchor=north east, font=\Huge] at (rel axis cs:0.98, 0.9) {\ce{ClO3}};
\node[draw=none, anchor=north east, font=\Huge] at (rel axis cs:0.98, 0.65) {$\mathrm{N_g}:7$};

\nextgroupplot[
tick align=outside,
tick pos=left,
x grid style={darkgray176},
xmin=0.8, xmax=42.5,
xticklabel style = {font=\huge},
yticklabel style = {font=\huge},
xtick distance=4,
yticklabels={},
xtick distance=5,
xtick style={color=black},
ymin=-0.044461245, ymax=1.049736145,
mark size=5pt,
width=\textwidth,
height=0.4\textwidth,
legend style={draw=none, at={(0.4,0.45)},fill=none, font=\Huge},
]

\addplot+[draw=blue, dash pattern=on 4pt off 2pt, legend image post style={sharp plot, dash pattern=on 4pt off 2pt}, mark options={fill=blue}]
table{%
x  y
0	0.9999999
1	0.9999999
2	0.9999181
3	0.999914
4	0.9998996
5	0.99989945
6	0.99988575
7	0.99988085
8	0.9998739
9	0.99987255
10	0.99975185
11	0.99974875
12	0.99289055
13	0.9925748
14	0.9925506
15	0.99190425
16	0.98995845
17	0.98968415
18	0.9876928
19	0.9852163
20	0.9780217
21	0.9718851
22	0.93243865
23	0.9135787
24	0.9123673
25	0.0873163
26	0.08607625
27	0.06728955
28	0.0099126
29	0.00900915
30	0.00900805
31	0.0076444
32	0.0076024
33	0.00686975
34	0.005706
35	0.00545815
36	0.00545265
37	0.00534365
38	0.00527265
39	0.0051744
40	0.0051619
};
\addplot+[draw=brown, mark=triangle*, mark options={fill=brown}, legend image post style={sharp plot}]
table {%
0	0.99999995
1	0.9999999
2	0.99991865
3	0.99989915
4	0.9998985
5	0.999896
6	0.99988515
7	0.99988225
8	0.9998779
9	0.99987205
10	0.9997519
11	0.99974515
12	0.99329855
13	0.9932943
14	0.992679
15	0.9922565
16	0.9914579
17	0.98927785
18	0.98912315
19	0.98234635
20	0.97402135
21	0.9722598
22	0.93344335
23	0.9134667
24	0.91022
25	0.08954745
26	0.08627905
27	0.0658338
28	0.00920735
29	0.00892565
30	0.008547
31	0.00753125
32	0.0072742
33	0.0071088
34	0.0067717
35	0.00647865
36	0.00574065
37	0.00564575
38	0.00553225
39	0.00534495
40	0.0051272
};
\legend{Softmax, Trigonometric}
\node[draw=none, anchor=north east, font=\Huge] at (rel axis cs:0.98, 0.9) {\ce{ClOOCl}};
\node[draw=none, anchor=north east, font=\Huge] at (rel axis cs:0.98, 0.65) {$\mathrm{N_g}:5$};

\nextgroupplot[
tick align=outside,
tick pos=left,
x grid style={darkgray176},
xmin=0.8, xmax=28.5,
xticklabel style = {font=\huge},
yticklabel style = {font=\huge},
xtick distance=4,
yticklabels={},
xtick style={color=black},
ymin=-0.0433123325, ymax=1.0496813825,
mark size=5pt,
width=\textwidth,
height=0.4\textwidth,
legend style={draw=none, at={(0.4,0.45)},fill=none, font=\Huge},
]
\addplot+[draw=blue, dash pattern=on 4pt off 2pt, legend image post style={sharp plot, dash pattern=on 4pt off 2pt}, mark options={fill=blue}]
table{%
x  y
0	0.99999985
1	0.9999486
2	0.9998949
3	0.99989385
4	0.99985975
5	0.99980445
6	0.99972995
7	0.99052675
8	0.9895075
9	0.9892861
10	0.98886725
11	0.98804815
12	0.97847305
13	0.97418945
14	0.9702821
15	0.75124495
16	0.5
17	0.24863105
18	0.02431545
19	0.0128591
20	0.011835
21	0.00901665
22	0.00801225
23	0.00712565
24	0.00696875
25	0.00695595
26	0.00680135
27	0.0061153
};
\addplot+[draw=brown, mark=triangle*, mark options={fill=brown}, legend image post style={sharp plot}]
table {%
0	0.9999999
1	0.99990125
2	0.99989425
3	0.9998911
4	0.99988295
5	0.9998408
6	0.9997633
7	0.989827
8	0.9896187
9	0.98951365
10	0.9892758
11	0.9873062
12	0.97947535
13	0.9743997
14	0.9686357
15	0.75079715
16	0.5
17	0.2490701
18	0.0240886
19	0.0135759
20	0.01253625
21	0.00872815
22	0.0072835
23	0.00710215
24	0.00703745
25	0.00698335
26	0.0068489
27	0.00543745
};
\legend{Softmax, Trigonometric}
\node[draw=none, anchor=north east, font=\Huge] at (rel axis cs:0.98, 0.9) {\ce{ClOO}};
\node[draw=none, anchor=north east, font=\Huge] at (rel axis cs:0.98, 0.65) {$\mathrm{N_g}:6$};

\draw (axis cs:0.95,0.9) node[
  scale=0.6,
  fill=white,
  draw=none,
  line width=0.4pt,
  inner sep=3.3pt,
  fill opacity=0.5,
  anchor=north east,
  text=black,
  rotate=0.0
]{$ClOO$};

\nextgroupplot[
tick align=outside,
tick pos=left,
x grid style={darkgray176},
xmin=0.8, xmax=16,
xticklabel style = {font=\huge},
yticklabel style = {font=\huge},
xtick distance=3,
xtick style={color=black},
y grid style={darkgray176},
ylabel={Occupation \\ Number},
ylabel style={align=center, yshift=15pt, font=\huge},
ymin=-0.04466947, ymax=1.04967147,
ytick style={color=black},
mark size=5pt,
width=\textwidth,
height=0.4\textwidth,
legend style={draw=none, at={(0.4,0.45)},fill=none, font=\Huge},
]

\addplot+[draw=blue, dash pattern=on 4pt off 2pt, legend image post style={sharp plot, dash pattern=on 4pt off 2pt}, mark options={fill=blue}]
table{%
x  y
0	0.99992905
1	0.99992835
2	0.9999018
3	0.9942285
4	0.99370495
5	0.99369915
6	0.993631
7	0.9935482
8	0.99342935
9	0.99296705
10	0.9916508
11	0.91224135
12	0.911466
13	0.08849465
14	0.0876904
15	0.0050532
};
\addplot+[draw=brown, mark=triangle*, mark options={fill=brown}, legend image post style={sharp plot}]
table {%
0	0.9999293
1	0.9999286
2	0.99990195
3	0.9937504
4	0.9936924
5	0.99367905
6	0.9936349
7	0.99362565
8	0.99351395
9	0.99318205
10	0.99158205
11	0.91175045
12	0.9115547
13	0.08838685
14	0.0881887
15	0.00511875
};
\legend{Softmax, Trigonometric}
\node[draw=none, anchor=north east, font=\Huge] at (rel axis cs:0.98, 0.9) {\ce{F2O}};
\node[draw=none, anchor=north east, font=\Huge] at (rel axis cs:0.98, 0.65) {$\mathrm{N_g}:8$};

\nextgroupplot[
tick align=outside,
tick pos=left,
x grid style={darkgray176},
xmin=0.8, xmax=22.2,
xticklabel style = {font=\huge},
yticklabel style = {font=\huge},
xtick distance=4,
yticklabels={},
xtick style={color=black},
ymin=-0.044380415, ymax=1.049655715,
mark size=5pt,
width=\textwidth,
height=0.4\textwidth,
legend style={draw=none, at={(0.4,0.45)},fill=none, font=\Huge},
]

\addplot+[draw=blue, dash pattern=on 4pt off 2pt, legend image post style={sharp plot, dash pattern=on 4pt off 2pt}, mark options={fill=blue}]
table{%
x  y
0	0.99992705
1	0.99989595
2	0.9998953
3	0.99239605
4	0.99232925
5	0.99229575
6	0.99034675
7	0.9901013
8	0.98984735
9	0.9891797
10	0.9768321
11	0.7949093
12	0.5
13	0.2050389
14	0.02167245
15	0.00709975
16	0.0068509
17	0.00674205
18	0.00621225
19	0.00526785
20	0.005215
21	0.0051901
};
\addplot+[draw=brown, mark=triangle*, mark options={fill=brown}, legend image post style={sharp plot}]
table {%
0	0.99992705
1	0.99989515
2	0.99989475
3	0.99237685
4	0.9923651
5	0.99225975
6	0.99041295
7	0.9899545
8	0.98959415
9	0.98925445
10	0.9769281
11	0.7948724
12	0.5
13	0.2050776
14	0.02158145
15	0.007195
16	0.00703545
17	0.00664705
18	0.00616115
19	0.00525595
20	0.0052285
21	0.005186
};
\legend{Softmax, Trigonometric}
\node[draw=none, anchor=north east, font=\Huge] at (rel axis cs:0.98, 0.9) {\ce{FO2}};
\node[draw=none, anchor=north east, font=\Huge] at (rel axis cs:0.98, 0.65) {$\mathrm{N_g}:8$};

\nextgroupplot[
tick align=outside,
tick pos=left,
x grid style={darkgray176},
xmin=0.8, xmax=22.1,
xticklabel style = {font=\huge},
yticklabel style = {font=\huge},
xtick distance=4,
yticklabels={},
xtick style={color=black},
ymin=-0.044661105, ymax=1.049670705,
mark size=5pt,
width=\textwidth,
height=0.4\textwidth,
legend style={draw=none, at={(0.4,0.45)},fill=none, font=\Huge},
]

\addplot+[draw=blue, dash pattern=on 4pt off 2pt, legend image post style={sharp plot, dash pattern=on 4pt off 2pt}, mark options={fill=blue}]
table{%
x  y
0	0.999928
1	0.999928
2	0.99990005
3	0.999899
4	0.99293775
5	0.99293665
6	0.99291655
7	0.99291265
8	0.9929077
9	0.9928743
10	0.992635
11	0.9924075
12	0.99226865
13	0.991952
14	0.97142095
15	0.8550207
16	0.8549526
17	0.14500925
18	0.144951
19	0.0274556
20	0.00533165
21	0.0053138
};
\addplot+[draw=brown, mark=triangle*, mark options={fill=brown}, legend image post style={sharp plot}]
table {%
0	0.99992885
1	0.9999281
2	0.9999003
3	0.99989925
4	0.99294515
5	0.99294195
6	0.9929244
7	0.9929098
8	0.9928963
9	0.992871
10	0.9926621
11	0.99244165
12	0.99218455
13	0.9919234
14	0.9715019
15	0.8550788
16	0.85499215
17	0.14498075
18	0.1448836
19	0.02736035
20	0.00535825
21	0.0053219
};
\legend{Softmax, Trigonometric}
\node[draw=none, anchor=north east, font=\Huge] at (rel axis cs:0.98, 0.9) {\ce{FOOF}};
\node[draw=none, anchor=north east, font=\Huge] at (rel axis cs:0.98, 0.65) {$\mathrm{N_g}:8$};

\nextgroupplot[
tick align=outside,
tick pos=left,
x grid style={darkgray176},
xmin=0.8, xmax=21.3,
xticklabel style = {font=\huge},
yticklabel style = {font=\huge},
xtick distance=3,
xtick style={color=black},
y grid style={darkgray176},
ylabel={Occupation \\ Number},
ylabel style={align=center, yshift=15pt, font=\huge},
ymin=-0.0446378, ymax=1.0496378,
ytick style={color=black},
mark size=5pt,
width=\textwidth,
height=0.4\textwidth,
legend style={draw=none, at={(0.4,0.45)},fill=none, font=\Huge},
]

\addplot+[draw=blue, dash pattern=on 4pt off 2pt, legend image post style={sharp plot, dash pattern=on 4pt off 2pt}, mark options={fill=blue}]
table{%
x  y
0	0.9998977
1	0.9998975
2	0.99989625
3	0.99250755
4	0.9910837
5	0.9909406
6	0.99049055
7	0.9904408
8	0.9899861
9	0.9642374
10	0.96412035
11	0.68743145
12	0.31254435
13	0.0347239
14	0.0345345
15	0.00626335
16	0.00609585
17	0.00605315
18	0.005866
19	0.0056283
20	0.005539
};
\addplot+[draw=brown, mark=triangle*, mark options={fill=brown}, legend image post style={sharp plot}]
table {%
0	0.9998974
1	0.9998971
2	0.99989625
3	0.9929818
4	0.9905804
5	0.99057575
6	0.9904521
7	0.99043775
8	0.99038555
9	0.96399135
10	0.96393665
11	0.68815925
12	0.31182185
13	0.03566935
14	0.0349028
15	0.00617755
16	0.00606105
17	0.00597
18	0.0058527
19	0.0058364
20	0.00530835
};
\legend{Softmax, Trigonometric}
\node[draw=none, anchor=north east, font=\Huge] at (rel axis cs:0.98, 0.9) {\ce{O3}};
\node[draw=none, anchor=north east, font=\Huge] at (rel axis cs:0.98, 0.65) {$\mathrm{N_g}:9$};

\nextgroupplot[
tick align=outside,
tick pos=left,
x grid style={darkgray176},
xmin=0.8, xmax=26.7,
xtick style={color=black},
ymin=-0.0446519775, ymax=1.0497452275,
xticklabel style = {font=\huge},
yticklabel style = {font=\huge},
xtick distance=5,
yticklabels={},
mark size=5pt,
width=\textwidth,
height=0.4\textwidth,
legend style={draw=none, at={(0.4,0.45)},fill=none, font=\Huge},
]

\addplot+[draw=blue, dash pattern=on 4pt off 2pt, legend image post style={sharp plot, dash pattern=on 4pt off 2pt}, mark options={fill=blue}]
table{%
x  y
0	0.9999999
1	0.999928
2	0.99989885
3	0.99989815
4	0.99984085
5	0.99981555
6	0.9998152
7	0.99135955
8	0.9905351
9	0.99036575
10	0.98931995
11	0.98722865
12	0.9864168
13	0.98583595
14	0.97636075
15	0.9725805
16	0.5
17	0.02709775
18	0.0203876
19	0.01140245
20	0.0111266
21	0.0081788
22	0.00741535
23	0.00729935
24	0.00712155
25	0.00588625
};
\addplot+[draw=brown, mark=triangle*, mark options={fill=brown}, legend image post style={sharp plot}]
table {%
0	0.9999999
1	0.99991835
2	0.99989975
3	0.99989895
4	0.9998322
5	0.99983145
6	0.99982485
7	0.99169985
8	0.99107475
9	0.99107465
10	0.99032635
11	0.98646825
12	0.98613485
13	0.97974585
14	0.9790367
15	0.97670625
16	0.5
17	0.01863465
18	0.01828365
19	0.0182352
20	0.01301975
21	0.00733595
22	0.0073316
23	0.0069429
24	0.00685525
25	0.00592295
};
\legend{Softmax, Trigonometric}
\node[draw=none, anchor=north east, font=\Huge] at (rel axis cs:0.98, 0.9) {\ce{OClO}};
\node[draw=none, anchor=north east, font=\Huge] at (rel axis cs:0.98, 0.65) {$\mathrm{N_g}:6$};

\nextgroupplot[
tick align=outside,
tick pos=left,
x grid style={darkgray176},
xmin=0.8, xmax=11.5,
xtick style={color=black},
ymin=-0.04463506, ymax=1.04966936,
xticklabel style = {font=\huge},
yticklabel style = {font=\huge},
xtick distance=2,
yticklabels={},
mark size=5pt,
width=\textwidth,
height=0.4\textwidth,
legend style={draw=none, at={(0.4,0.45)},fill=none, font=\Huge},
]

\addplot+[draw=blue, dash pattern=on 4pt off 2pt, legend image post style={sharp plot, dash pattern=on 4pt off 2pt}, mark options={fill=blue}]
table{%
x  y
0	0.99992805
1	0.999898
2	0.99385465
3	0.9933833
4	0.99312165
5	0.98984885
6	0.98860985
7	0.92594505
8	0.5
9	0.073872
10	0.00686225
11	0.0059329
};
\addplot+[draw=brown, mark=triangle*, mark options={fill=brown}, legend image post style={sharp plot}]
table {%
0	0.99992815
1	0.99989875
2	0.99389055
3	0.9935635
4	0.99288905
5	0.98966595
6	0.9890934
7	0.9259402
8	0.5
9	0.0739038
10	0.00671355
11	0.00616225
};
\legend{Softmax, Trigonometric}
\node[draw=none, anchor=north east, font=\Huge] at (rel axis cs:0.98, 0.9) {\ce{OF}};
\node[draw=none, anchor=north east, font=\Huge] at (rel axis cs:0.98, 0.65) {$\mathrm{N_g}:8$};

\draw (axis cs:0.95,0.9) node[
  scale=0.6,
  fill=white,
  draw=none,
  line width=0.4pt,
  inner sep=3.3pt,
  fill opacity=0.5,
  anchor=north east,
  text=black,
  rotate=0.0
]{$OF$};

\nextgroupplot[
tick align=outside,
tick pos=left,
x grid style={darkgray176},
xlabel={Relative Natural Orbital Number},
xlabel style={align=center,yshift=-10pt, font=\huge},
xticklabel style = {font=\huge},
xmin=0.8, xmax=38,
xticklabel style = {font=\huge},
yticklabel style = {font=\huge},
xtick distance=5,
xtick style={color=black},
y grid style={darkgray176},
ylabel={Occupation \\ Number},
ylabel style={align=center, yshift=15pt, font=\huge},
yticklabel style = {font=\huge},
ymin=-0.0444011325, ymax=1.0497332825,
ytick style={color=black},
mark size=5pt,
width=\textwidth,
height=0.4\textwidth,
legend style={draw=none, at={(0.4,0.45)},fill=none, font=\Huge},
]

\addplot+[draw=blue, dash pattern=on 4pt off 2pt, legend image post style={sharp plot, dash pattern=on 4pt off 2pt}, mark options={fill=blue}]
table{%
x  y
0	0.99999995
1	0.9999999
2	0.9999999
3	0.99985015
4	0.9998455
5	0.99981985
6	0.99981075
7	0.99980955
8	0.999805
9	0.99979875
10	0.999787
11	0.9997599
12	0.9997551
13	0.99975285
14	0.99974275
15	0.992213
16	0.99160055
17	0.9795698
18	0.9687506
19	0.96787235
20	0.96473655
21	0.96379245
22	0.95817935
23	0.70105395
24	0.29858805
25	0.03754495
26	0.02254685
27	0.020729
28	0.01293525
29	0.01216085
30	0.01023405
31	0.009997
32	0.00996085
33	0.0094055
34	0.00890415
35	0.00693465
36	0.00519195
};
\addplot+[draw=brown, mark=triangle*, mark options={fill=brown}, legend image post style={sharp plot}]
table {%
0	0.99999995
1	0.9999999
2	0.9999999
3	0.9998455
4	0.99982165
5	0.99981705
6	0.9998167
7	0.99980215
8	0.99979475
9	0.999788
10	0.9997863
11	0.9997862
12	0.99978195
13	0.99975185
14	0.9997452
15	0.99183265
16	0.99180045
17	0.97897565
18	0.9643802
19	0.9641525
20	0.96209125
21	0.96196585
22	0.95514535
23	0.69594095
24	0.30367875
25	0.03007835
26	0.02984055
27	0.025124
28	0.01536655
29	0.01294815
30	0.0129255
31	0.01074565
32	0.0107128
33	0.0102435
34	0.01013075
35	0.0087513
36	0.007112
};
\legend{Softmax, Trigonometric}
\node[draw=none, anchor=north east, font=\Huge] at (rel axis cs:0.98, 0.9) {\ce{S3}};
\node[draw=none, anchor=north east, font=\Huge] at (rel axis cs:0.98, 0.65) {$\mathrm{N_g}:4$};

\nextgroupplot[
tick align=outside,
tick pos=left,
x grid style={darkgray176},
xlabel={Relative Natural Orbital Number},
xlabel style={align=center,yshift=-10pt, font=\huge},
xmin=0.8, xmax=51,
xticklabel style = {font=\huge},
yticklabel style = {font=\huge},
xtick distance=5,
yticklabels={},
xtick style={color=black},
ymin=-0.0444725875, ymax=1.0497367375,
mark size=5pt,
width=\textwidth,
height=0.4\textwidth,
legend style={draw=none, at={(0.4,0.45)},fill=none, font=\Huge},
]

\addplot+[draw=blue, dash pattern=on 4pt off 2pt, legend image post style={sharp plot, dash pattern=on 4pt off 2pt}, mark options={fill=blue}]
table{%
x  y
0	0.99999995
1	0.99999995
2	0.9999999
3	0.9999999
4	0.9998562
5	0.9998525
6	0.9998241
7	0.99981295
8	0.9998056
9	0.99980355
10	0.9997917
11	0.9997902
12	0.9997803
13	0.9997749
14	0.9997663
15	0.99975665
16	0.99974615
17	0.99974165
18	0.9997287
19	0.99971475
20	0.99222985
21	0.99084025
22	0.98486735
23	0.9786029
24	0.96502545
25	0.96207075
26	0.96145555
27	0.95755215
28	0.95610975
29	0.94958045
30	0.91079755
31	0.60417875
32	0.3953964
33	0.0882369
34	0.04597355
35	0.032572
36	0.0217371
37	0.0194458
38	0.01823905
39	0.01743915
40	0.01342895
41	0.01158315
42	0.01107825
43	0.01070955
44	0.01010065
45	0.00957285
46	0.0094659
47	0.00564255
48	0.0052615
};
\addplot+[draw=brown, mark=triangle*, mark options={fill=brown}, legend image post style={sharp plot}]
table {%
0	0.99999995
1	0.99999995
2	0.9999999
3	0.9999999
4	0.9998562
5	0.9998525
6	0.9998241
7	0.99981295
8	0.9998056
9	0.99980355
10	0.9997917
11	0.9997902
12	0.9997803
13	0.9997749
14	0.9997663
15	0.99975665
16	0.99974615
17	0.99974165
18	0.9997287
19	0.99971475
20	0.99222985
21	0.99084025
22	0.98486735
23	0.9786029
24	0.96502545
25	0.96207075
26	0.96145555
27	0.95755215
28	0.95610975
29	0.94958045
30	0.91079755
31	0.60417875
32	0.3953964
33	0.0882369
34	0.04597355
35	0.032572
36	0.0217371
37	0.0194458
38	0.01823905
39	0.01743915
40	0.01342895
41	0.01158315
42	0.01107825
43	0.01070955
44	0.01010065
45	0.00957285
46	0.0094659
47	0.00564255
48	0.0052615
};
\legend{Softmax, Trigonometric}
\node[draw=none, anchor=north east, font=\Huge] at (rel axis cs:0.98, 0.9) {\ce{S4}};
\node[draw=none, anchor=north east, font=\Huge] at (rel axis cs:0.98, 0.65) {$\mathrm{N_g}:4$};

\end{groupplot}

\end{tikzpicture}


%% file: Figures/W4-17-MR-dE-M-PNOF7_new.tex

\begin{tikzpicture}[scale=0.5]
\definecolor{darkgray176}{RGB}{176,176,176}

\begin{groupplot}[group style={group size=2 by 1,vertical sep=2cm}]
\nextgroupplot[
colorbar,
colorbar horizontal,
colorbar style={xlabel={M-Diagnostic}, 
point meta min=0, point meta max=1,
xtick={0, 0.2, 0.4, 0.6, 0.8, 1.}, 
font=\Huge, tick label style={font=\LARGE}, at={(0.15,-0.2)}, width=0.7\textwidth},
colormap={mymap}{[1pt]
 rgb(0pt)=(0,0.135112,0.304751);
  rgb(1pt)=(0,0.138068,0.311105);
  rgb(2pt)=(0,0.141013,0.317579);
  rgb(3pt)=(0,0.143951,0.323982);
  rgb(4pt)=(0,0.146877,0.330479);
  rgb(5pt)=(0,0.149791,0.337065);
  rgb(6pt)=(0,0.152673,0.343704);
  rgb(7pt)=(0,0.155377,0.3505);
  rgb(8pt)=(0,0.157932,0.357521);
  rgb(9pt)=(0,0.160495,0.364534);
  rgb(10pt)=(0,0.163058,0.371608);
  rgb(11pt)=(0,0.165621,0.378769);
  rgb(12pt)=(0,0.168204,0.385902);
  rgb(13pt)=(0,0.1708,0.3931);
  rgb(14pt)=(0,0.17342,0.400353);
  rgb(15pt)=(0,0.176082,0.407577);
  rgb(16pt)=(0,0.178802,0.414764);
  rgb(17pt)=(0,0.18161,0.421859);
  rgb(18pt)=(0,0.18455,0.428802);
  rgb(19pt)=(0,0.186915,0.435532);
  rgb(20pt)=(0,0.188769,0.439563);
  rgb(21pt)=(0,0.19095,0.441085);
  rgb(22pt)=(0,0.193366,0.441561);
  rgb(23pt)=(0.003602,0.195911,0.441564);
  rgb(24pt)=(0.017852,0.198528,0.441248);
  rgb(25pt)=(0.03211,0.201199,0.440785);
  rgb(26pt)=(0.046205,0.203903,0.440196);
  rgb(27pt)=(0.058378,0.206629,0.439531);
  rgb(28pt)=(0.068968,0.209372,0.438863);
  rgb(29pt)=(0.078624,0.212122,0.438105);
  rgb(30pt)=(0.087465,0.214879,0.437342);
  rgb(31pt)=(0.095645,0.217643,0.436593);
  rgb(32pt)=(0.103401,0.220406,0.43579);
  rgb(33pt)=(0.110658,0.22317,0.435067);
  rgb(34pt)=(0.117612,0.225935,0.434308);
  rgb(35pt)=(0.124291,0.228697,0.433547);
  rgb(36pt)=(0.130669,0.231458,0.43284);
  rgb(37pt)=(0.13683,0.234216,0.432148);
  rgb(38pt)=(0.142852,0.236972,0.431404);
  rgb(39pt)=(0.148638,0.239724,0.430752);
  rgb(40pt)=(0.154261,0.242475,0.43012);
  rgb(41pt)=(0.159733,0.245221,0.429528);
  rgb(42pt)=(0.165113,0.247965,0.428908);
  rgb(43pt)=(0.170362,0.250707,0.428325);
  rgb(44pt)=(0.17549,0.253444,0.42779);
  rgb(45pt)=(0.180503,0.25618,0.427299);
  rgb(46pt)=(0.185453,0.258914,0.426788);
  rgb(47pt)=(0.190303,0.261644,0.426329);
  rgb(48pt)=(0.195057,0.264372,0.425924);
  rgb(49pt)=(0.199764,0.267099,0.425497);
  rgb(50pt)=(0.204385,0.269823,0.425126);
  rgb(51pt)=(0.208926,0.272546,0.424809);
  rgb(52pt)=(0.213431,0.275266,0.42448);
  rgb(53pt)=(0.217863,0.277985,0.424206);
  rgb(54pt)=(0.222264,0.280702,0.423914);
  rgb(55pt)=(0.226598,0.283419,0.423678);
  rgb(56pt)=(0.230871,0.286134,0.423498);
  rgb(57pt)=(0.23512,0.288848,0.423304);
  rgb(58pt)=(0.239312,0.291562,0.423167);
  rgb(59pt)=(0.243485,0.294274,0.423014);
  rgb(60pt)=(0.247605,0.296986,0.422917);
  rgb(61pt)=(0.251675,0.299698,0.422873);
  rgb(62pt)=(0.255731,0.302409,0.422814);
  rgb(63pt)=(0.25974,0.30512,0.42281);
  rgb(64pt)=(0.263738,0.307831,0.422789);
  rgb(65pt)=(0.267693,0.310542,0.422821);
  rgb(66pt)=(0.271639,0.313253,0.422837);
  rgb(67pt)=(0.275513,0.315965,0.422979);
  rgb(68pt)=(0.279411,0.318677,0.423031);
  rgb(69pt)=(0.28324,0.32139,0.423211);
  rgb(70pt)=(0.287065,0.324103,0.423373);
  rgb(71pt)=(0.290884,0.326816,0.423517);
  rgb(72pt)=(0.294669,0.329531,0.423716);
  rgb(73pt)=(0.298421,0.332247,0.423973);
  rgb(74pt)=(0.302169,0.334963,0.424213);
  rgb(75pt)=(0.305886,0.337681,0.424512);
  rgb(76pt)=(0.309601,0.340399,0.42479);
  rgb(77pt)=(0.313287,0.34312,0.42512);
  rgb(78pt)=(0.316941,0.345842,0.425512);
  rgb(79pt)=(0.320595,0.348565,0.425889);
  rgb(80pt)=(0.32425,0.351289,0.42625);
  rgb(81pt)=(0.327875,0.354016,0.42667);
  rgb(82pt)=(0.331474,0.356744,0.427144);
  rgb(83pt)=(0.335073,0.359474,0.427605);
  rgb(84pt)=(0.338673,0.362206,0.428053);
  rgb(85pt)=(0.342246,0.364939,0.428559);
  rgb(86pt)=(0.345793,0.367676,0.429127);
  rgb(87pt)=(0.349341,0.370414,0.429685);
  rgb(88pt)=(0.352892,0.373153,0.430226);
  rgb(89pt)=(0.356418,0.375896,0.430823);
  rgb(90pt)=(0.359916,0.378641,0.431501);
  rgb(91pt)=(0.363446,0.381388,0.432075);
  rgb(92pt)=(0.366923,0.384139,0.432796);
  rgb(93pt)=(0.37043,0.38689,0.433428);
  rgb(94pt)=(0.373884,0.389646,0.434209);
  rgb(95pt)=(0.377371,0.392404,0.43489);
  rgb(96pt)=(0.38083,0.395164,0.435653);
  rgb(97pt)=(0.384268,0.397928,0.436475);
  rgb(98pt)=(0.387705,0.400694,0.437305);
  rgb(99pt)=(0.391151,0.403464,0.438096);
  rgb(100pt)=(0.394568,0.406236,0.438986);
  rgb(101pt)=(0.397991,0.409011,0.439848);
  rgb(102pt)=(0.401418,0.41179,0.440708);
  rgb(103pt)=(0.40482,0.414572,0.441642);
  rgb(104pt)=(0.408226,0.417357,0.44257);
  rgb(105pt)=(0.411607,0.420145,0.443577);
  rgb(106pt)=(0.414992,0.422937,0.444578);
  rgb(107pt)=(0.418383,0.425733,0.44556);
  rgb(108pt)=(0.421748,0.428531,0.44664);
  rgb(109pt)=(0.42512,0.431334,0.447692);
  rgb(110pt)=(0.428462,0.43414,0.448864);
  rgb(111pt)=(0.431817,0.43695,0.449982);
  rgb(112pt)=(0.435168,0.439763,0.451134);
  rgb(113pt)=(0.438504,0.44258,0.452341);
  rgb(114pt)=(0.44181,0.445402,0.453659);
  rgb(115pt)=(0.445148,0.448226,0.454885);
  rgb(116pt)=(0.448447,0.451053,0.456264);
  rgb(117pt)=(0.451759,0.453887,0.457582);
  rgb(118pt)=(0.455072,0.456718,0.458976);
  rgb(119pt)=(0.458366,0.459552,0.460457);
  rgb(120pt)=(0.461616,0.462405,0.461969);
  rgb(121pt)=(0.464947,0.465241,0.463395);
  rgb(122pt)=(0.468254,0.468083,0.464908);
  rgb(123pt)=(0.471501,0.47096,0.466357);
  rgb(124pt)=(0.474812,0.473832,0.467681);
  rgb(125pt)=(0.478186,0.476699,0.468845);
  rgb(126pt)=(0.481622,0.479573,0.469767);
  rgb(127pt)=(0.485141,0.482451,0.470384);
  rgb(128pt)=(0.488697,0.485318,0.471008);
  rgb(129pt)=(0.492278,0.488198,0.471453);
  rgb(130pt)=(0.495913,0.491076,0.471751);
  rgb(131pt)=(0.499552,0.49396,0.472032);
  rgb(132pt)=(0.503185,0.496851,0.472305);
  rgb(133pt)=(0.506866,0.499743,0.472432);
  rgb(134pt)=(0.51054,0.502643,0.47255);
  rgb(135pt)=(0.514226,0.505546,0.47264);
  rgb(136pt)=(0.51792,0.508454,0.472707);
  rgb(137pt)=(0.521643,0.511367,0.472639);
  rgb(138pt)=(0.525348,0.514285,0.47266);
  rgb(139pt)=(0.529086,0.517207,0.472543);
  rgb(140pt)=(0.532829,0.520135,0.472401);
  rgb(141pt)=(0.536553,0.523067,0.472352);
  rgb(142pt)=(0.540307,0.526005,0.472163);
  rgb(143pt)=(0.544069,0.528948,0.471947);
  rgb(144pt)=(0.54784,0.531895,0.471704);
  rgb(145pt)=(0.551612,0.534849,0.471439);
  rgb(146pt)=(0.555393,0.537807,0.471147);
  rgb(147pt)=(0.559181,0.540771,0.470829);
  rgb(148pt)=(0.562972,0.543741,0.470488);
  rgb(149pt)=(0.566802,0.546715,0.469988);
  rgb(150pt)=(0.570607,0.549695,0.469593);
  rgb(151pt)=(0.574417,0.552682,0.469172);
  rgb(152pt)=(0.578236,0.555673,0.468724);
  rgb(153pt)=(0.582087,0.55867,0.468118);
  rgb(154pt)=(0.585916,0.561674,0.467618);
  rgb(155pt)=(0.589753,0.564682,0.46709);
  rgb(156pt)=(0.593622,0.567697,0.466401);
  rgb(157pt)=(0.597469,0.570718,0.465821);
  rgb(158pt)=(0.601354,0.573743,0.465074);
  rgb(159pt)=(0.605211,0.576777,0.464441);
  rgb(160pt)=(0.609105,0.579816,0.463638);
  rgb(161pt)=(0.612977,0.582861,0.46295);
  rgb(162pt)=(0.616852,0.585913,0.462237);
  rgb(163pt)=(0.620765,0.58897,0.461351);
  rgb(164pt)=(0.624654,0.592034,0.460583);
  rgb(165pt)=(0.628576,0.595104,0.459641);
  rgb(166pt)=(0.632506,0.59818,0.458668);
  rgb(167pt)=(0.636412,0.601264,0.457818);
  rgb(168pt)=(0.640352,0.604354,0.456791);
  rgb(169pt)=(0.64427,0.60745,0.455886);
  rgb(170pt)=(0.648222,0.610553,0.454801);
  rgb(171pt)=(0.652178,0.613664,0.453689);
  rgb(172pt)=(0.656114,0.61678,0.452702);
  rgb(173pt)=(0.660082,0.619904,0.451534);
  rgb(174pt)=(0.664055,0.623034,0.450338);
  rgb(175pt)=(0.668008,0.626171,0.44927);
  rgb(176pt)=(0.671991,0.629316,0.448018);
  rgb(177pt)=(0.675981,0.632468,0.446736);
  rgb(178pt)=(0.679979,0.635626,0.445424);
  rgb(179pt)=(0.68395,0.638793,0.444251);
  rgb(180pt)=(0.687957,0.641966,0.442886);
  rgb(181pt)=(0.691971,0.645145,0.441491);
  rgb(182pt)=(0.695985,0.648334,0.440072);
  rgb(183pt)=(0.700008,0.651529,0.438624);
  rgb(184pt)=(0.704037,0.654731,0.437147);
  rgb(185pt)=(0.708067,0.657942,0.435647);
  rgb(186pt)=(0.712105,0.66116,0.434117);
  rgb(187pt)=(0.716177,0.664384,0.432386);
  rgb(188pt)=(0.720222,0.667618,0.430805);
  rgb(189pt)=(0.724274,0.670859,0.429194);
  rgb(190pt)=(0.728334,0.674107,0.427554);
  rgb(191pt)=(0.732422,0.677364,0.425717);
  rgb(192pt)=(0.736488,0.680629,0.424028);
  rgb(193pt)=(0.740589,0.6839,0.422131);
  rgb(194pt)=(0.744664,0.687181,0.420393);
  rgb(195pt)=(0.748772,0.69047,0.418448);
  rgb(196pt)=(0.752886,0.693766,0.416472);
  rgb(197pt)=(0.756975,0.697071,0.414659);
  rgb(198pt)=(0.761096,0.700384,0.412638);
  rgb(199pt)=(0.765223,0.703705,0.410587);
  rgb(200pt)=(0.769353,0.707035,0.408516);
  rgb(201pt)=(0.773486,0.710373,0.406422);
  rgb(202pt)=(0.777651,0.713719,0.404112);
  rgb(203pt)=(0.781795,0.717074,0.401966);
  rgb(204pt)=(0.785965,0.720438,0.399613);
  rgb(205pt)=(0.790116,0.72381,0.397423);
  rgb(206pt)=(0.794298,0.72719,0.395016);
  rgb(207pt)=(0.79848,0.73058,0.392597);
  rgb(208pt)=(0.802667,0.733978,0.390153);
  rgb(209pt)=(0.806859,0.737385,0.387684);
  rgb(210pt)=(0.811054,0.740801,0.385198);
  rgb(211pt)=(0.815274,0.744226,0.382504);
  rgb(212pt)=(0.819499,0.747659,0.379785);
  rgb(213pt)=(0.823729,0.751101,0.377043);
  rgb(214pt)=(0.827959,0.754553,0.374292);
  rgb(215pt)=(0.832192,0.758014,0.371529);
  rgb(216pt)=(0.836429,0.761483,0.368747);
  rgb(217pt)=(0.840693,0.764962,0.365746);
  rgb(218pt)=(0.844957,0.76845,0.362741);
  rgb(219pt)=(0.849223,0.771947,0.359729);
  rgb(220pt)=(0.853515,0.775454,0.3565);
  rgb(221pt)=(0.857809,0.778969,0.353259);
  rgb(222pt)=(0.862105,0.782494,0.350011);
  rgb(223pt)=(0.866421,0.786028,0.346571);
  rgb(224pt)=(0.870717,0.789572,0.343333);
  rgb(225pt)=(0.875057,0.793125,0.339685);
  rgb(226pt)=(0.879378,0.796687,0.336241);
  rgb(227pt)=(0.88372,0.800258,0.332599);
  rgb(228pt)=(0.888081,0.803839,0.32877);
  rgb(229pt)=(0.89244,0.80743,0.324968);
  rgb(230pt)=(0.896818,0.81103,0.320982);
  rgb(231pt)=(0.901195,0.814639,0.317021);
  rgb(232pt)=(0.905589,0.818257,0.312889);
  rgb(233pt)=(0.91,0.821885,0.308594);
  rgb(234pt)=(0.914407,0.825522,0.304348);
  rgb(235pt)=(0.918828,0.829168,0.29996);
  rgb(236pt)=(0.923279,0.832822,0.295244);
  rgb(237pt)=(0.927724,0.836486,0.290611);
  rgb(238pt)=(0.93218,0.840159,0.28588);
  rgb(239pt)=(0.93666,0.843841,0.280876);
  rgb(240pt)=(0.941147,0.84753,0.275815);
  rgb(241pt)=(0.945654,0.851228,0.270532);
  rgb(242pt)=(0.950178,0.854933,0.265085);
  rgb(243pt)=(0.954725,0.858646,0.259365);
  rgb(244pt)=(0.959284,0.862365,0.253563);
  rgb(245pt)=(0.963872,0.866089,0.247445);
  rgb(246pt)=(0.968469,0.869819,0.24131);
  rgb(247pt)=(0.973114,0.87355,0.234677);
  rgb(248pt)=(0.97778,0.877281,0.227954);
  rgb(249pt)=(0.982497,0.881008,0.220878);
  rgb(250pt)=(0.987293,0.884718,0.213336);
  rgb(251pt)=(0.992218,0.888385,0.205468);
  rgb(252pt)=(0.994847,0.892954,0.203445);
  rgb(253pt)=(0.995249,0.898384,0.207561);
  rgb(254pt)=(0.995503,0.903866,0.21237);
  rgb(255pt)=(0.995737,0.909344,0.217772)
},
point meta max=1,
point meta min=0.05,
tick align=outside,
tick pos=left,
x grid style={darkgray176},
xlabel={\(\displaystyle R_\text{T:S} \)},
xmin=-9, xmax=45,
xlabel style={font=\Huge},
ytick={-0.1,1.1,2.3,3.5,4.7,5.9,7.1,8.3,9.5,10.7,11.9,13.1,14.3,15.5,16.7,17.9,19.1},
yticklabels={
  \(\displaystyle \ce{B_2}\),
  \(\displaystyle \ce{BN}\),
  \(\displaystyle \ce{C_2}\),
  \(\displaystyle \ce{Cl_2O}\),
  \(\displaystyle \ce{ClF_3}\),
  \(\displaystyle \ce{ClF_5}\),
  \(\displaystyle \ce{ClO_3}\),
  \(\displaystyle \ce{ClOOCl}\),
  \(\displaystyle \ce{ClOO}\),
  \(\displaystyle \ce{F_2O}\),
  \(\displaystyle \ce{FO_2}\),
  \(\displaystyle \ce{FOOF}\),
  \(\displaystyle \ce{O_3}\),
  \(\displaystyle \ce{OClO}\),
  \(\displaystyle \ce{OF}\),
  \(\displaystyle \ce{S_3}\),
  \(\displaystyle \ce{S_4}\)
},
tick label style={font=\LARGE},
y grid style={darkgray176},
ymin=-0.8, ymax=20,
ytick style={color=black},
ytick distance=5, 
width=0.45\textwidth,
height=\textwidth,
ylabel style={align=center, inner sep=0pt, font=\LARGE\bfseries\boldmath},
xlabel style={align=center, inner sep=5pt, font=\LARGE\bfseries\boldmath},
x tick label style={font=\LARGE\bfseries\boldmath},
y tick label style={font=\LARGE\bfseries\boldmath},
]


\path [draw=black, dotted]
(axis cs:-25,-.1)
--(axis cs:45.770257,-.1);
\path [draw=black, dotted]
(axis cs:-25,1.1)
--(axis cs:45.770257,1.1);
\path [draw=black, dotted]
(axis cs:-25, 2.3)
--(axis cs:45.770257, 2.3);
\path [draw=black, dotted]
(axis cs:-25, 3.5)
--(axis cs:45.770257, 3.5);
\path [draw=black, dotted]
(axis cs:-25, 4.7)
--(axis cs:45.770257, 4.7);
\path [draw=black, dotted]
(axis cs:-25, 5.9)
--(axis cs:45.770257, 5.9);
\path [draw=black, dotted]
(axis cs:-25, 7.1)
--(axis cs:45.770257, 7.1);
\path [draw=black, dotted]
(axis cs:-25, 8.3)
--(axis cs:45.770257, 8.3);
\path [draw=black, dotted]
(axis cs:-25, 9.5)
--(axis cs:45.770257, 9.5);
\path [draw=black, dotted]
(axis cs:-25, 10.7)
--(axis cs:45.770257, 10.7);
\path [draw=black, dotted]
(axis cs:-25, 11.9)
--(axis cs:45.770257, 11.9);
\path [draw=black, dotted]
(axis cs:-25, 13,1)
--(axis cs:45.770257, 13.1);
\path [draw=black, dotted]
(axis cs:-25, 14.3)
--(axis cs:45.770257, 14.3);
\path [draw=black, dotted]
(axis cs:-25, 15.5)
--(axis cs:45.770257, 15.5);
\path [draw=black, dotted]
(axis cs:-25, 16.7)
--(axis cs:45.770257, 16.7);
\path [draw=black, dotted]
(axis cs:-25, 17.9)
--(axis cs:45.770257, 17.9);
\path [draw=black, dotted]
(axis cs:-25, 19.1)
--(axis cs:45.770257, 19.1);
\addplot [
  colormap={mymap}{[1pt]addplot
 rgb(0pt)=(0,0.135112,0.304751);
  rgb(1pt)=(0,0.138068,0.311105);
  rgb(2pt)=(0,0.141013,0.317579);
  rgb(3pt)=(0,0.143951,0.323982);
  rgb(4pt)=(0,0.146877,0.330479);
  rgb(5pt)=(0,0.149791,0.337065);
  rgb(6pt)=(0,0.152673,0.343704);
  rgb(7pt)=(0,0.155377,0.3505);
  rgb(8pt)=(0,0.157932,0.357521);
  rgb(9pt)=(0,0.160495,0.364534);
  rgb(10pt)=(0,0.163058,0.371608);
  rgb(11pt)=(0,0.165621,0.378769);
  rgb(12pt)=(0,0.168204,0.385902);
  rgb(13pt)=(0,0.1708,0.3931);
  rgb(14pt)=(0,0.17342,0.400353);
  rgb(15pt)=(0,0.176082,0.407577);
  rgb(16pt)=(0,0.178802,0.414764);
  rgb(17pt)=(0,0.18161,0.421859);
  rgb(18pt)=(0,0.18455,0.428802);
  rgb(19pt)=(0,0.186915,0.435532);
  rgb(20pt)=(0,0.188769,0.439563);
  rgb(21pt)=(0,0.19095,0.441085);
  rgb(22pt)=(0,0.193366,0.441561);
  rgb(23pt)=(0.003602,0.195911,0.441564);
  rgb(24pt)=(0.017852,0.198528,0.441248);
  rgb(25pt)=(0.03211,0.201199,0.440785);
  rgb(26pt)=(0.046205,0.203903,0.440196);
  rgb(27pt)=(0.058378,0.206629,0.439531);
  rgb(28pt)=(0.068968,0.209372,0.438863);
  rgb(29pt)=(0.078624,0.212122,0.438105);
  rgb(30pt)=(0.087465,0.214879,0.437342);
  rgb(31pt)=(0.095645,0.217643,0.436593);
  rgb(32pt)=(0.103401,0.220406,0.43579);
  rgb(33pt)=(0.110658,0.22317,0.435067);
  rgb(34pt)=(0.117612,0.225935,0.434308);
  rgb(35pt)=(0.124291,0.228697,0.433547);
  rgb(36pt)=(0.130669,0.231458,0.43284);
  rgb(37pt)=(0.13683,0.234216,0.432148);
  rgb(38pt)=(0.142852,0.236972,0.431404);
  rgb(39pt)=(0.148638,0.239724,0.430752);
  rgb(40pt)=(0.154261,0.242475,0.43012);
  rgb(41pt)=(0.159733,0.245221,0.429528);
  rgb(42pt)=(0.165113,0.247965,0.428908);
  rgb(43pt)=(0.170362,0.250707,0.428325);
  rgb(44pt)=(0.17549,0.253444,0.42779);
  rgb(45pt)=(0.180503,0.25618,0.427299);
  rgb(46pt)=(0.185453,0.258914,0.426788);
  rgb(47pt)=(0.190303,0.261644,0.426329);
  rgb(48pt)=(0.195057,0.264372,0.425924);
  rgb(49pt)=(0.199764,0.267099,0.425497);
  rgb(50pt)=(0.204385,0.269823,0.425126);
  rgb(51pt)=(0.208926,0.272546,0.424809);
  rgb(52pt)=(0.213431,0.275266,0.42448);
  rgb(53pt)=(0.217863,0.277985,0.424206);
  rgb(54pt)=(0.222264,0.280702,0.423914);
  rgb(55pt)=(0.226598,0.283419,0.423678);
  rgb(56pt)=(0.230871,0.286134,0.423498);
  rgb(57pt)=(0.23512,0.288848,0.423304);
  rgb(58pt)=(0.239312,0.291562,0.423167);
  rgb(59pt)=(0.243485,0.294274,0.423014);
  rgb(60pt)=(0.247605,0.296986,0.422917);
  rgb(61pt)=(0.251675,0.299698,0.422873);
  rgb(62pt)=(0.255731,0.302409,0.422814);
  rgb(63pt)=(0.25974,0.30512,0.42281);
  rgb(64pt)=(0.263738,0.307831,0.422789);
  rgb(65pt)=(0.267693,0.310542,0.422821);
  rgb(66pt)=(0.271639,0.313253,0.422837);
  rgb(67pt)=(0.275513,0.315965,0.422979);
  rgb(68pt)=(0.279411,0.318677,0.423031);
  rgb(69pt)=(0.28324,0.32139,0.423211);
  rgb(70pt)=(0.287065,0.324103,0.423373);
  rgb(71pt)=(0.290884,0.326816,0.423517);
  rgb(72pt)=(0.294669,0.329531,0.423716);
  rgb(73pt)=(0.298421,0.332247,0.423973);
  rgb(74pt)=(0.302169,0.334963,0.424213);
  rgb(75pt)=(0.305886,0.337681,0.424512);
  rgb(76pt)=(0.309601,0.340399,0.42479);
  rgb(77pt)=(0.313287,0.34312,0.42512);
  rgb(78pt)=(0.316941,0.345842,0.425512);
  rgb(79pt)=(0.320595,0.348565,0.425889);
  rgb(80pt)=(0.32425,0.351289,0.42625);
  rgb(81pt)=(0.327875,0.354016,0.42667);
  rgb(82pt)=(0.331474,0.356744,0.427144);
  rgb(83pt)=(0.335073,0.359474,0.427605);
  rgb(84pt)=(0.338673,0.362206,0.428053);
  rgb(85pt)=(0.342246,0.364939,0.428559);
  rgb(86pt)=(0.345793,0.367676,0.429127);
  rgb(87pt)=(0.349341,0.370414,0.429685);
  rgb(88pt)=(0.352892,0.373153,0.430226);
  rgb(89pt)=(0.356418,0.375896,0.430823);
  rgb(90pt)=(0.359916,0.378641,0.431501);
  rgb(91pt)=(0.363446,0.381388,0.432075);
  rgb(92pt)=(0.366923,0.384139,0.432796);
  rgb(93pt)=(0.37043,0.38689,0.433428);
  rgb(94pt)=(0.373884,0.389646,0.434209);
  rgb(95pt)=(0.377371,0.392404,0.43489);
  rgb(96pt)=(0.38083,0.395164,0.435653);
  rgb(97pt)=(0.384268,0.397928,0.436475);
  rgb(98pt)=(0.387705,0.400694,0.437305);
  rgb(99pt)=(0.391151,0.403464,0.438096);
  rgb(100pt)=(0.394568,0.406236,0.438986);
  rgb(101pt)=(0.397991,0.409011,0.439848);
  rgb(102pt)=(0.401418,0.41179,0.440708);
  rgb(103pt)=(0.40482,0.414572,0.441642);
  rgb(104pt)=(0.408226,0.417357,0.44257);
  rgb(105pt)=(0.411607,0.420145,0.443577);
  rgb(106pt)=(0.414992,0.422937,0.444578);
  rgb(107pt)=(0.418383,0.425733,0.44556);
  rgb(108pt)=(0.421748,0.428531,0.44664);
  rgb(109pt)=(0.42512,0.431334,0.447692);
  rgb(110pt)=(0.428462,0.43414,0.448864);
  rgb(111pt)=(0.431817,0.43695,0.449982);
  rgb(112pt)=(0.435168,0.439763,0.451134);
  rgb(113pt)=(0.438504,0.44258,0.452341);
  rgb(114pt)=(0.44181,0.445402,0.453659);
  rgb(115pt)=(0.445148,0.448226,0.454885);
  rgb(116pt)=(0.448447,0.451053,0.456264);
  rgb(117pt)=(0.451759,0.453887,0.457582);
  rgb(118pt)=(0.455072,0.456718,0.458976);
  rgb(119pt)=(0.458366,0.459552,0.460457);
  rgb(120pt)=(0.461616,0.462405,0.461969);
  rgb(121pt)=(0.464947,0.465241,0.463395);
  rgb(122pt)=(0.468254,0.468083,0.464908);
  rgb(123pt)=(0.471501,0.47096,0.466357);
  rgb(124pt)=(0.474812,0.473832,0.467681);
  rgb(125pt)=(0.478186,0.476699,0.468845);
  rgb(126pt)=(0.481622,0.479573,0.469767);
  rgb(127pt)=(0.485141,0.482451,0.470384);
  rgb(128pt)=(0.488697,0.485318,0.471008);
  rgb(129pt)=(0.492278,0.488198,0.471453);
  rgb(130pt)=(0.495913,0.491076,0.471751);
  rgb(131pt)=(0.499552,0.49396,0.472032);
  rgb(132pt)=(0.503185,0.496851,0.472305);
  rgb(133pt)=(0.506866,0.499743,0.472432);
  rgb(134pt)=(0.51054,0.502643,0.47255);
  rgb(135pt)=(0.514226,0.505546,0.47264);
  rgb(136pt)=(0.51792,0.508454,0.472707);
  rgb(137pt)=(0.521643,0.511367,0.472639);
  rgb(138pt)=(0.525348,0.514285,0.47266);
  rgb(139pt)=(0.529086,0.517207,0.472543);
  rgb(140pt)=(0.532829,0.520135,0.472401);
  rgb(141pt)=(0.536553,0.523067,0.472352);
  rgb(142pt)=(0.540307,0.526005,0.472163);
  rgb(143pt)=(0.544069,0.528948,0.471947);
  rgb(144pt)=(0.54784,0.531895,0.471704);
  rgb(145pt)=(0.551612,0.534849,0.471439);
  rgb(146pt)=(0.555393,0.537807,0.471147);
  rgb(147pt)=(0.559181,0.540771,0.470829);
  rgb(148pt)=(0.562972,0.543741,0.470488);
  rgb(149pt)=(0.566802,0.546715,0.469988);
  rgb(150pt)=(0.570607,0.549695,0.469593);
  rgb(151pt)=(0.574417,0.552682,0.469172);
  rgb(152pt)=(0.578236,0.555673,0.468724);
  rgb(153pt)=(0.582087,0.55867,0.468118);
  rgb(154pt)=(0.585916,0.561674,0.467618);
  rgb(155pt)=(0.589753,0.564682,0.46709);
  rgb(156pt)=(0.593622,0.567697,0.466401);
  rgb(157pt)=(0.597469,0.570718,0.465821);
  rgb(158pt)=(0.601354,0.573743,0.465074);
  rgb(159pt)=(0.605211,0.576777,0.464441);
  rgb(160pt)=(0.609105,0.579816,0.463638);
  rgb(161pt)=(0.612977,0.582861,0.46295);
  rgb(162pt)=(0.616852,0.585913,0.462237);
  rgb(163pt)=(0.620765,0.58897,0.461351);
  rgb(164pt)=(0.624654,0.592034,0.460583);
  rgb(165pt)=(0.628576,0.595104,0.459641);
  rgb(166pt)=(0.632506,0.59818,0.458668);
  rgb(167pt)=(0.636412,0.601264,0.457818);
  rgb(168pt)=(0.640352,0.604354,0.456791);
  rgb(169pt)=(0.64427,0.60745,0.455886);
  rgb(170pt)=(0.648222,0.610553,0.454801);
  rgb(171pt)=(0.652178,0.613664,0.453689);
  rgb(172pt)=(0.656114,0.61678,0.452702);
  rgb(173pt)=(0.660082,0.619904,0.451534);
  rgb(174pt)=(0.664055,0.623034,0.450338);
  rgb(175pt)=(0.668008,0.626171,0.44927);
  rgb(176pt)=(0.671991,0.629316,0.448018);
  rgb(177pt)=(0.675981,0.632468,0.446736);
  rgb(178pt)=(0.679979,0.635626,0.445424);
  rgb(179pt)=(0.68395,0.638793,0.444251);
  rgb(180pt)=(0.687957,0.641966,0.442886);
  rgb(181pt)=(0.691971,0.645145,0.441491);
  rgb(182pt)=(0.695985,0.648334,0.440072);
  rgb(183pt)=(0.700008,0.651529,0.438624);
  rgb(184pt)=(0.704037,0.654731,0.437147);
  rgb(185pt)=(0.708067,0.657942,0.435647);
  rgb(186pt)=(0.712105,0.66116,0.434117);
  rgb(187pt)=(0.716177,0.664384,0.432386);
  rgb(188pt)=(0.720222,0.667618,0.430805);
  rgb(189pt)=(0.724274,0.670859,0.429194);
  rgb(190pt)=(0.728334,0.674107,0.427554);
  rgb(191pt)=(0.732422,0.677364,0.425717);
  rgb(192pt)=(0.736488,0.680629,0.424028);
  rgb(193pt)=(0.740589,0.6839,0.422131);
  rgb(194pt)=(0.744664,0.687181,0.420393);
  rgb(195pt)=(0.748772,0.69047,0.418448);
  rgb(196pt)=(0.752886,0.693766,0.416472);
  rgb(197pt)=(0.756975,0.697071,0.414659);
  rgb(198pt)=(0.761096,0.700384,0.412638);
  rgb(199pt)=(0.765223,0.703705,0.410587);
  rgb(200pt)=(0.769353,0.707035,0.408516);
  rgb(201pt)=(0.773486,0.710373,0.406422);
  rgb(202pt)=(0.777651,0.713719,0.404112);
  rgb(203pt)=(0.781795,0.717074,0.401966);
  rgb(204pt)=(0.785965,0.720438,0.399613);
  rgb(205pt)=(0.790116,0.72381,0.397423);
  rgb(206pt)=(0.794298,0.72719,0.395016);
  rgb(207pt)=(0.79848,0.73058,0.392597);
  rgb(208pt)=(0.802667,0.733978,0.390153);
  rgb(209pt)=(0.806859,0.737385,0.387684);
  rgb(210pt)=(0.811054,0.740801,0.385198);
  rgb(211pt)=(0.815274,0.744226,0.382504);
  rgb(212pt)=(0.819499,0.747659,0.379785);
  rgb(213pt)=(0.823729,0.751101,0.377043);
  rgb(214pt)=(0.827959,0.754553,0.374292);
  rgb(215pt)=(0.832192,0.758014,0.371529);
  rgb(216pt)=(0.836429,0.761483,0.368747);
  rgb(217pt)=(0.840693,0.764962,0.365746);
  rgb(218pt)=(0.844957,0.76845,0.362741);
  rgb(219pt)=(0.849223,0.771947,0.359729);
  rgb(220pt)=(0.853515,0.775454,0.3565);
  rgb(221pt)=(0.857809,0.778969,0.353259);
  rgb(222pt)=(0.862105,0.782494,0.350011);
  rgb(223pt)=(0.866421,0.786028,0.346571);
  rgb(224pt)=(0.870717,0.789572,0.343333);
  rgb(225pt)=(0.875057,0.793125,0.339685);
  rgb(226pt)=(0.879378,0.796687,0.336241);
  rgb(227pt)=(0.88372,0.800258,0.332599);
  rgb(228pt)=(0.888081,0.803839,0.32877);
  rgb(229pt)=(0.89244,0.80743,0.324968);
  rgb(230pt)=(0.896818,0.81103,0.320982);
  rgb(231pt)=(0.901195,0.814639,0.317021);
  rgb(232pt)=(0.905589,0.818257,0.312889);
  rgb(233pt)=(0.91,0.821885,0.308594);
  rgb(234pt)=(0.914407,0.825522,0.304348);
  rgb(235pt)=(0.918828,0.829168,0.29996);
  rgb(236pt)=(0.923279,0.832822,0.295244);
  rgb(237pt)=(0.927724,0.836486,0.290611);
  rgb(238pt)=(0.93218,0.840159,0.28588);
  rgb(239pt)=(0.93666,0.843841,0.280876);
  rgb(240pt)=(0.941147,0.84753,0.275815);
  rgb(241pt)=(0.945654,0.851228,0.270532);
  rgb(242pt)=(0.950178,0.854933,0.265085);
  rgb(243pt)=(0.954725,0.858646,0.259365);
  rgb(244pt)=(0.959284,0.862365,0.253563);
  rgb(245pt)=(0.963872,0.866089,0.247445);
  rgb(246pt)=(0.968469,0.869819,0.24131);
  rgb(247pt)=(0.973114,0.87355,0.234677);
  rgb(248pt)=(0.97778,0.877281,0.227954);
  rgb(249pt)=(0.982497,0.881008,0.220878);
  rgb(250pt)=(0.987293,0.884718,0.213336);
  rgb(251pt)=(0.992218,0.888385,0.205468);
  rgb(252pt)=(0.994847,0.892954,0.203445);
  rgb(253pt)=(0.995249,0.898384,0.207561);
  rgb(254pt)=(0.995503,0.903866,0.21237);
  rgb(255pt)=(0.995737,0.909344,0.217772)
},
  only marks,
  mark size = 8pt, 
  scatter,
  scatter src=explicit
]
table [x=x, y=y, meta=colordata]{%
x  y  colordata
4.828326	-0.1	0.57
8.467836	1.1	1
2.423077	2.3	0.99
14.345291	3.5	0.16
25.398551	4.7	0.18
25.846154	5.9	0.09
10.553846	7.1	0.05
19.474576	8.3	0.18
14.820809	9.5	0.5
10.457516	10.7	0.18
7.291480	11.9	0.41
18.037313	13.1	0.29
4.024691    14.3	0.63
29.192661	15.5	0.08
10.275362	16.7	0.15
8.864078	17.9	0.6
11.529101   19.1	0.79
};
\node[draw=none, anchor=north east, font=\Huge] at (rel axis cs:0.27, 0.965) {\bf(a)};

\hfill

\nextgroupplot[
point meta max=1,
point meta min=0.05,
tick align=outside,
tick pos=left,
x grid style={darkgray176},
xlabel={$\displaystyle \Delta {E}  / 10^{-2}$ \\ (kcal/mol)},
xmin=-2, xmax=2,
xlabel style={font=\Huge},
xtick style={color=black},
tick label style={font=\LARGE},
y grid style={darkgray176},
ymin=-0.8, ymax=20, 
yticklabels={},
ytick style={draw=none},
width=0.45\textwidth,
height=\textwidth,
ylabel style={align=center, inner sep=0pt, font=\LARGE\bfseries\boldmath},
xlabel style={align=center, inner sep=5pt, font=\LARGE\bfseries\boldmath},
x tick label style={font=\LARGE\bfseries\boldmath},
y tick label style={font=\LARGE\bfseries\boldmath},
]
\path [draw=black, semithick]
(axis cs:0,-0.8)
--(axis cs:0,20);

\path [draw=black, dotted]
(axis cs:-25,-.1)
--(axis cs:6.5,-.1);
\path [draw=black, dotted]
(axis cs:-25,1.1)
--(axis cs:6.5,1.1);
\path [draw=black, dotted]
(axis cs:-25, 2.3)
--(axis cs:6.5, 2.3);
\path [draw=black, dotted]
(axis cs:-25, 3.5)
--(axis cs:6.5, 3.5);
\path [draw=black, dotted]
(axis cs:-25, 4.7)
--(axis cs:6.5, 4.7);
\path [draw=black, dotted]
(axis cs:-25, 5.9)
--(axis cs:6.5, 5.9);
\path [draw=black, dotted]
(axis cs:-25, 7.1)
--(axis cs:6.5, 7.1);
\path [draw=black, dotted]
(axis cs:-25, 8.3)
--(axis cs:6.5, 8.3);
\path [draw=black, dotted]
(axis cs:-25, 9.5)
--(axis cs:6.5, 9.5);
\path [draw=black, dotted]
(axis cs:-25, 10.7)
--(axis cs:6.5, 10.7);
\path [draw=black, dotted]
(axis cs:-25, 11.9)
--(axis cs:6.5, 11.9);
\path [draw=black, dotted]
(axis cs:-25, 13,1)
--(axis cs:6.5, 13.1);
\path [draw=black, dotted]
(axis cs:-25, 14.3)
--(axis cs:6.5, 14.3);
\path [draw=black, dotted]
(axis cs:-25, 15.5)
--(axis cs:6.5, 15.5);
\path [draw=black, dotted]
(axis cs:-25, 16.7)
--(axis cs:6.5, 16.7);
\path [draw=black, dotted]
(axis cs:-25, 17.9)
--(axis cs:6.5, 17.9);
\path [draw=black, dotted]
(axis cs:-25, 19.1)
--(axis cs:6.5, 19.1);
\addplot [
  colormap={mymap}{[1pt]addplot
 rgb(0pt)=(0,0.135112,0.304751);
  rgb(1pt)=(0,0.138068,0.311105);
  rgb(2pt)=(0,0.141013,0.317579);
  rgb(3pt)=(0,0.143951,0.323982);
  rgb(4pt)=(0,0.146877,0.330479);
  rgb(5pt)=(0,0.149791,0.337065);
  rgb(6pt)=(0,0.152673,0.343704);
  rgb(7pt)=(0,0.155377,0.3505);
  rgb(8pt)=(0,0.157932,0.357521);
  rgb(9pt)=(0,0.160495,0.364534);
  rgb(10pt)=(0,0.163058,0.371608);
  rgb(11pt)=(0,0.165621,0.378769);
  rgb(12pt)=(0,0.168204,0.385902);
  rgb(13pt)=(0,0.1708,0.3931);
  rgb(14pt)=(0,0.17342,0.400353);
  rgb(15pt)=(0,0.176082,0.407577);
  rgb(16pt)=(0,0.178802,0.414764);
  rgb(17pt)=(0,0.18161,0.421859);
  rgb(18pt)=(0,0.18455,0.428802);
  rgb(19pt)=(0,0.186915,0.435532);
  rgb(20pt)=(0,0.188769,0.439563);
  rgb(21pt)=(0,0.19095,0.441085);
  rgb(22pt)=(0,0.193366,0.441561);
  rgb(23pt)=(0.003602,0.195911,0.441564);
  rgb(24pt)=(0.017852,0.198528,0.441248);
  rgb(25pt)=(0.03211,0.201199,0.440785);
  rgb(26pt)=(0.046205,0.203903,0.440196);
  rgb(27pt)=(0.058378,0.206629,0.439531);
  rgb(28pt)=(0.068968,0.209372,0.438863);
  rgb(29pt)=(0.078624,0.212122,0.438105);
  rgb(30pt)=(0.087465,0.214879,0.437342);
  rgb(31pt)=(0.095645,0.217643,0.436593);
  rgb(32pt)=(0.103401,0.220406,0.43579);
  rgb(33pt)=(0.110658,0.22317,0.435067);
  rgb(34pt)=(0.117612,0.225935,0.434308);
  rgb(35pt)=(0.124291,0.228697,0.433547);
  rgb(36pt)=(0.130669,0.231458,0.43284);
  rgb(37pt)=(0.13683,0.234216,0.432148);
  rgb(38pt)=(0.142852,0.236972,0.431404);
  rgb(39pt)=(0.148638,0.239724,0.430752);
  rgb(40pt)=(0.154261,0.242475,0.43012);
  rgb(41pt)=(0.159733,0.245221,0.429528);
  rgb(42pt)=(0.165113,0.247965,0.428908);
  rgb(43pt)=(0.170362,0.250707,0.428325);
  rgb(44pt)=(0.17549,0.253444,0.42779);
  rgb(45pt)=(0.180503,0.25618,0.427299);
  rgb(46pt)=(0.185453,0.258914,0.426788);
  rgb(47pt)=(0.190303,0.261644,0.426329);
  rgb(48pt)=(0.195057,0.264372,0.425924);
  rgb(49pt)=(0.199764,0.267099,0.425497);
  rgb(50pt)=(0.204385,0.269823,0.425126);
  rgb(51pt)=(0.208926,0.272546,0.424809);
  rgb(52pt)=(0.213431,0.275266,0.42448);
  rgb(53pt)=(0.217863,0.277985,0.424206);
  rgb(54pt)=(0.222264,0.280702,0.423914);
  rgb(55pt)=(0.226598,0.283419,0.423678);
  rgb(56pt)=(0.230871,0.286134,0.423498);
  rgb(57pt)=(0.23512,0.288848,0.423304);
  rgb(58pt)=(0.239312,0.291562,0.423167);
  rgb(59pt)=(0.243485,0.294274,0.423014);
  rgb(60pt)=(0.247605,0.296986,0.422917);
  rgb(61pt)=(0.251675,0.299698,0.422873);
  rgb(62pt)=(0.255731,0.302409,0.422814);
  rgb(63pt)=(0.25974,0.30512,0.42281);
  rgb(64pt)=(0.263738,0.307831,0.422789);
  rgb(65pt)=(0.267693,0.310542,0.422821);
  rgb(66pt)=(0.271639,0.313253,0.422837);
  rgb(67pt)=(0.275513,0.315965,0.422979);
  rgb(68pt)=(0.279411,0.318677,0.423031);
  rgb(69pt)=(0.28324,0.32139,0.423211);
  rgb(70pt)=(0.287065,0.324103,0.423373);
  rgb(71pt)=(0.290884,0.326816,0.423517);
  rgb(72pt)=(0.294669,0.329531,0.423716);
  rgb(73pt)=(0.298421,0.332247,0.423973);
  rgb(74pt)=(0.302169,0.334963,0.424213);
  rgb(75pt)=(0.305886,0.337681,0.424512);
  rgb(76pt)=(0.309601,0.340399,0.42479);
  rgb(77pt)=(0.313287,0.34312,0.42512);
  rgb(78pt)=(0.316941,0.345842,0.425512);
  rgb(79pt)=(0.320595,0.348565,0.425889);
  rgb(80pt)=(0.32425,0.351289,0.42625);
  rgb(81pt)=(0.327875,0.354016,0.42667);
  rgb(82pt)=(0.331474,0.356744,0.427144);
  rgb(83pt)=(0.335073,0.359474,0.427605);
  rgb(84pt)=(0.338673,0.362206,0.428053);
  rgb(85pt)=(0.342246,0.364939,0.428559);
  rgb(86pt)=(0.345793,0.367676,0.429127);
  rgb(87pt)=(0.349341,0.370414,0.429685);
  rgb(88pt)=(0.352892,0.373153,0.430226);
  rgb(89pt)=(0.356418,0.375896,0.430823);
  rgb(90pt)=(0.359916,0.378641,0.431501);
  rgb(91pt)=(0.363446,0.381388,0.432075);
  rgb(92pt)=(0.366923,0.384139,0.432796);
  rgb(93pt)=(0.37043,0.38689,0.433428);
  rgb(94pt)=(0.373884,0.389646,0.434209);
  rgb(95pt)=(0.377371,0.392404,0.43489);
  rgb(96pt)=(0.38083,0.395164,0.435653);
  rgb(97pt)=(0.384268,0.397928,0.436475);
  rgb(98pt)=(0.387705,0.400694,0.437305);
  rgb(99pt)=(0.391151,0.403464,0.438096);
  rgb(100pt)=(0.394568,0.406236,0.438986);
  rgb(101pt)=(0.397991,0.409011,0.439848);
  rgb(102pt)=(0.401418,0.41179,0.440708);
  rgb(103pt)=(0.40482,0.414572,0.441642);
  rgb(104pt)=(0.408226,0.417357,0.44257);
  rgb(105pt)=(0.411607,0.420145,0.443577);
  rgb(106pt)=(0.414992,0.422937,0.444578);
  rgb(107pt)=(0.418383,0.425733,0.44556);
  rgb(108pt)=(0.421748,0.428531,0.44664);
  rgb(109pt)=(0.42512,0.431334,0.447692);
  rgb(110pt)=(0.428462,0.43414,0.448864);
  rgb(111pt)=(0.431817,0.43695,0.449982);
  rgb(112pt)=(0.435168,0.439763,0.451134);
  rgb(113pt)=(0.438504,0.44258,0.452341);
  rgb(114pt)=(0.44181,0.445402,0.453659);
  rgb(115pt)=(0.445148,0.448226,0.454885);
  rgb(116pt)=(0.448447,0.451053,0.456264);
  rgb(117pt)=(0.451759,0.453887,0.457582);
  rgb(118pt)=(0.455072,0.456718,0.458976);
  rgb(119pt)=(0.458366,0.459552,0.460457);
  rgb(120pt)=(0.461616,0.462405,0.461969);
  rgb(121pt)=(0.464947,0.465241,0.463395);
  rgb(122pt)=(0.468254,0.468083,0.464908);
  rgb(123pt)=(0.471501,0.47096,0.466357);
  rgb(124pt)=(0.474812,0.473832,0.467681);
  rgb(125pt)=(0.478186,0.476699,0.468845);
  rgb(126pt)=(0.481622,0.479573,0.469767);
  rgb(127pt)=(0.485141,0.482451,0.470384);
  rgb(128pt)=(0.488697,0.485318,0.471008);
  rgb(129pt)=(0.492278,0.488198,0.471453);
  rgb(130pt)=(0.495913,0.491076,0.471751);
  rgb(131pt)=(0.499552,0.49396,0.472032);
  rgb(132pt)=(0.503185,0.496851,0.472305);
  rgb(133pt)=(0.506866,0.499743,0.472432);
  rgb(134pt)=(0.51054,0.502643,0.47255);
  rgb(135pt)=(0.514226,0.505546,0.47264);
  rgb(136pt)=(0.51792,0.508454,0.472707);
  rgb(137pt)=(0.521643,0.511367,0.472639);
  rgb(138pt)=(0.525348,0.514285,0.47266);
  rgb(139pt)=(0.529086,0.517207,0.472543);
  rgb(140pt)=(0.532829,0.520135,0.472401);
  rgb(141pt)=(0.536553,0.523067,0.472352);
  rgb(142pt)=(0.540307,0.526005,0.472163);
  rgb(143pt)=(0.544069,0.528948,0.471947);
  rgb(144pt)=(0.54784,0.531895,0.471704);
  rgb(145pt)=(0.551612,0.534849,0.471439);
  rgb(146pt)=(0.555393,0.537807,0.471147);
  rgb(147pt)=(0.559181,0.540771,0.470829);
  rgb(148pt)=(0.562972,0.543741,0.470488);
  rgb(149pt)=(0.566802,0.546715,0.469988);
  rgb(150pt)=(0.570607,0.549695,0.469593);
  rgb(151pt)=(0.574417,0.552682,0.469172);
  rgb(152pt)=(0.578236,0.555673,0.468724);
  rgb(153pt)=(0.582087,0.55867,0.468118);
  rgb(154pt)=(0.585916,0.561674,0.467618);
  rgb(155pt)=(0.589753,0.564682,0.46709);
  rgb(156pt)=(0.593622,0.567697,0.466401);
  rgb(157pt)=(0.597469,0.570718,0.465821);
  rgb(158pt)=(0.601354,0.573743,0.465074);
  rgb(159pt)=(0.605211,0.576777,0.464441);
  rgb(160pt)=(0.609105,0.579816,0.463638);
  rgb(161pt)=(0.612977,0.582861,0.46295);
  rgb(162pt)=(0.616852,0.585913,0.462237);
  rgb(163pt)=(0.620765,0.58897,0.461351);
  rgb(164pt)=(0.624654,0.592034,0.460583);
  rgb(165pt)=(0.628576,0.595104,0.459641);
  rgb(166pt)=(0.632506,0.59818,0.458668);
  rgb(167pt)=(0.636412,0.601264,0.457818);
  rgb(168pt)=(0.640352,0.604354,0.456791);
  rgb(169pt)=(0.64427,0.60745,0.455886);
  rgb(170pt)=(0.648222,0.610553,0.454801);
  rgb(171pt)=(0.652178,0.613664,0.453689);
  rgb(172pt)=(0.656114,0.61678,0.452702);
  rgb(173pt)=(0.660082,0.619904,0.451534);
  rgb(174pt)=(0.664055,0.623034,0.450338);
  rgb(175pt)=(0.668008,0.626171,0.44927);
  rgb(176pt)=(0.671991,0.629316,0.448018);
  rgb(177pt)=(0.675981,0.632468,0.446736);
  rgb(178pt)=(0.679979,0.635626,0.445424);
  rgb(179pt)=(0.68395,0.638793,0.444251);
  rgb(180pt)=(0.687957,0.641966,0.442886);
  rgb(181pt)=(0.691971,0.645145,0.441491);
  rgb(182pt)=(0.695985,0.648334,0.440072);
  rgb(183pt)=(0.700008,0.651529,0.438624);
  rgb(184pt)=(0.704037,0.654731,0.437147);
  rgb(185pt)=(0.708067,0.657942,0.435647);
  rgb(186pt)=(0.712105,0.66116,0.434117);
  rgb(187pt)=(0.716177,0.664384,0.432386);
  rgb(188pt)=(0.720222,0.667618,0.430805);
  rgb(189pt)=(0.724274,0.670859,0.429194);
  rgb(190pt)=(0.728334,0.674107,0.427554);
  rgb(191pt)=(0.732422,0.677364,0.425717);
  rgb(192pt)=(0.736488,0.680629,0.424028);
  rgb(193pt)=(0.740589,0.6839,0.422131);
  rgb(194pt)=(0.744664,0.687181,0.420393);
  rgb(195pt)=(0.748772,0.69047,0.418448);
  rgb(196pt)=(0.752886,0.693766,0.416472);
  rgb(197pt)=(0.756975,0.697071,0.414659);
  rgb(198pt)=(0.761096,0.700384,0.412638);
  rgb(199pt)=(0.765223,0.703705,0.410587);
  rgb(200pt)=(0.769353,0.707035,0.408516);
  rgb(201pt)=(0.773486,0.710373,0.406422);
  rgb(202pt)=(0.777651,0.713719,0.404112);
  rgb(203pt)=(0.781795,0.717074,0.401966);
  rgb(204pt)=(0.785965,0.720438,0.399613);
  rgb(205pt)=(0.790116,0.72381,0.397423);
  rgb(206pt)=(0.794298,0.72719,0.395016);
  rgb(207pt)=(0.79848,0.73058,0.392597);
  rgb(208pt)=(0.802667,0.733978,0.390153);
  rgb(209pt)=(0.806859,0.737385,0.387684);
  rgb(210pt)=(0.811054,0.740801,0.385198);
  rgb(211pt)=(0.815274,0.744226,0.382504);
  rgb(212pt)=(0.819499,0.747659,0.379785);
  rgb(213pt)=(0.823729,0.751101,0.377043);
  rgb(214pt)=(0.827959,0.754553,0.374292);
  rgb(215pt)=(0.832192,0.758014,0.371529);
  rgb(216pt)=(0.836429,0.761483,0.368747);
  rgb(217pt)=(0.840693,0.764962,0.365746);
  rgb(218pt)=(0.844957,0.76845,0.362741);
  rgb(219pt)=(0.849223,0.771947,0.359729);
  rgb(220pt)=(0.853515,0.775454,0.3565);
  rgb(221pt)=(0.857809,0.778969,0.353259);
  rgb(222pt)=(0.862105,0.782494,0.350011);
  rgb(223pt)=(0.866421,0.786028,0.346571);
  rgb(224pt)=(0.870717,0.789572,0.343333);
  rgb(225pt)=(0.875057,0.793125,0.339685);
  rgb(226pt)=(0.879378,0.796687,0.336241);
  rgb(227pt)=(0.88372,0.800258,0.332599);
  rgb(228pt)=(0.888081,0.803839,0.32877);
  rgb(229pt)=(0.89244,0.80743,0.324968);
  rgb(230pt)=(0.896818,0.81103,0.320982);
  rgb(231pt)=(0.901195,0.814639,0.317021);
  rgb(232pt)=(0.905589,0.818257,0.312889);
  rgb(233pt)=(0.91,0.821885,0.308594);
  rgb(234pt)=(0.914407,0.825522,0.304348);
  rgb(235pt)=(0.918828,0.829168,0.29996);
  rgb(236pt)=(0.923279,0.832822,0.295244);
  rgb(237pt)=(0.927724,0.836486,0.290611);
  rgb(238pt)=(0.93218,0.840159,0.28588);
  rgb(239pt)=(0.93666,0.843841,0.280876);
  rgb(240pt)=(0.941147,0.84753,0.275815);
  rgb(241pt)=(0.945654,0.851228,0.270532);
  rgb(242pt)=(0.950178,0.854933,0.265085);
  rgb(243pt)=(0.954725,0.858646,0.259365);
  rgb(244pt)=(0.959284,0.862365,0.253563);
  rgb(245pt)=(0.963872,0.866089,0.247445);
  rgb(246pt)=(0.968469,0.869819,0.24131);
  rgb(247pt)=(0.973114,0.87355,0.234677);
  rgb(248pt)=(0.97778,0.877281,0.227954);
  rgb(249pt)=(0.982497,0.881008,0.220878);
  rgb(250pt)=(0.987293,0.884718,0.213336);
  rgb(251pt)=(0.992218,0.888385,0.205468);
  rgb(252pt)=(0.994847,0.892954,0.203445);
  rgb(253pt)=(0.995249,0.898384,0.207561);
  rgb(254pt)=(0.995503,0.903866,0.21237);
  rgb(255pt)=(0.995737,0.909344,0.217772)
},
  only marks,
  mark size = 8pt, 
  scatter,
  scatter src=explicit
]
table [x=x, y=y, meta=colordata]{%
x  y  colordata
 0.7064 -0.1 0.57
 0.5388  1.1 1.0
 0.5702  2.3 0.99
 0.5625  3.5 0.16
 0.5546  4.7 0.18
 0.0734  5.9 0.09
-0.0667  7.1 0.05
 0.4282  8.3 0.18
 0.3517  9.5 0.5
 0.9528  10.7 0.18
 0.2283  11.9 0.41
 1.5043  13.1 0.29
 0.5734  14.3 0.63
 0.1699  15.5 0.08
 0.3800  16.7 0.15
 0.1606  17.9 0.6
 0.4716  19.1 0.79
};
\node[draw=none, anchor=north east, font=\Huge] at (rel axis cs:0.28, 0.965) {\bf(b)};
\end{groupplot}
\end{tikzpicture}